\documentclass[preprint,12pt,authoryear]{elsarticle}
 
\usepackage{amsmath}
\usepackage{amssymb}
\usepackage{amsthm}
\usepackage{dsfont}

\usepackage{graphicx,subcaption}
\usepackage{array, tabularx, booktabs, multicol}
\usepackage{threeparttable}
\usepackage{rotating}
\usepackage{siunitx}
\usepackage{xcolor}
\usepackage[margin=1.2in]{geometry}
\usepackage[dvipsnames]{xcolor}

\usepackage{tikz}
\usetikzlibrary{arrows.meta, positioning, shapes.geometric, fit, backgrounds, calc}
\usepackage{pgfplots}
\pgfplotsset{compat=1.18}

\usepackage{algorithm}
\usepackage{algpseudocode}
\usepackage{bbm}

\usepackage{setspace}
\usepackage{comment}
\usepackage[hidelinks]{hyperref} 
\usepackage{multirow} 

\theoremstyle{definition}

\usepackage{natbib}

\usepackage{bm}

\definecolor{cData}{HTML}{E8EEF7}
\definecolor{cDataB}{HTML}{4A6FA5}
\definecolor{cMap}{HTML}{FCEEE3}
\definecolor{cMapB}{HTML}{C87941}
\definecolor{cState}{HTML}{E9F3EC}
\definecolor{cStateB}{HTML}{4E8A63}
\definecolor{cScore}{HTML}{F5E9F0}
\definecolor{cScoreB}{HTML}{9B5A82}
\definecolor{cGrey}{HTML}{F2F2F2}

\tikzset{
  data/.style   = {rectangle, rounded corners=2pt, draw=cDataB, thick,
                   fill=cData, minimum height=8mm, minimum width=15mm,
                   align=center, font=\small},
  map/.style    = {rectangle, rounded corners=2pt, draw=cMapB, thick,
                   fill=cMap, minimum height=9mm, minimum width=20mm,
                   align=center, font=\small},
  state/.style  = {rectangle, rounded corners=2pt, draw=cStateB, thick,
                   fill=cState, minimum height=8mm, minimum width=18mm,
                   align=center, font=\small},
  score/.style  = {rectangle, rounded corners=2pt, draw=cScoreB, thick,
                   fill=cScore, minimum height=8mm, minimum width=18mm,
                   align=center, font=\small},
  grp/.style    = {rectangle, rounded corners=3pt, draw=black!25,
                   dashed, fill=cGrey, inner sep=4mm},
  fl/.style     = {-{Latex[length=2.5mm]}, thick, draw=black!65},
  flb/.style    = {-{Latex[length=2.5mm]}, thick, draw=cStateB, dashed},
  lbl/.style    = {font=\scriptsize, inner sep=1.5pt, fill=white}
}

\journal{International Journal of Forecasting}

\begin{document}

\begin{frontmatter}

\title{Time-Varying Recalibration of Forecast Distributions}

\author[ucl]{Callum Wu\corref{cor1}}
\ead{callum.wu.24@ucl.ac.uk}

\author[ucl]{Jim E. Griffin}
\ead{j.griffin@ucl.ac.uk}

\cortext[cor1]{Corresponding author}

\affiliation[ucl]{organization={Department of Statistical Science, University College London},
            city={London},
            postcode={WC1E 6BT},
            country={United Kingdom}}

\begin{abstract}
Combined density forecasts are often miscalibrated, and the miscalibration can change over time. We propose a time-varying framework that recalibrates any combined forecast through applying a monotone map to its probability integral transform. We consider a Beta CDF or a rational quadratic spline as the map, and their parameters are updated either via score-driven filters combined with dynamic model averaging, or by a convolution-based update that additionally conditions on features of the forecast distribution. In simulations, the spline map corrects multimodal miscalibration that the Beta map cannot, and the convolution update achieves the highest log score when the miscalibration switches between regimes. We apply the method to nine G10 exchange rates and US inflation, recalibrating forecast distributions from both a loss-discounting linear pool and Bayesian predictive synthesis. In both applications, recalibration improves PIT uniformity, with gains in forecast accuracy most evident when the original forecasts are miscalibrated.
\end{abstract}


\end{frontmatter}

\section{Introduction}
\label{sec:intro}
Central banks, policy institutions and private forecasters now routinely publish probabilistic forecasts, because the full predictive distribution quantifies uncertainty about future outcomes. These forecasts are increasingly obtained by combining the outputs of several component models. Combination improves empirical performance and alleviates the risk of relying on a single misspecified model \citep{Bates1969,Clemen1989,Stock2004,Timmermann2006,Wang2023}. Early work combined point forecasts through weighted averages, with weights chosen to minimise mean squared error \citep{Bates1969,Clemen1986}. For probabilistic forecasts, the form of
combination depends on how each forecast is represented: by its cumulative distribution function (CDF), its density or its quantile function. A weighted average of the CDFs or the densities gives the linear pool \citep{Stone1961,Hall2007}. Averaging the quantile functions gives a quantile average, which is sharper than the linear pool and preserves the common shape of components from the same location-scale family \citep{Lichtendahl2013,Busetti2017}. Non-linear pools transform a weighted average of CDFs, as in the beta-transformed linear pool \citep{Ranjan2010,Gneiting2013} and the centered linear pools \citep{Knuppel2022}. 

Although many combination methods have been proposed, \citet{Aastveit2019} argue that most applications still focus on the linear pool. Suppose $N$ component models produce one-step-ahead predictive CDFs $F_{1,t},\dots,F_{N,t}$, with densities $f_{1,t},\dots,f_{N,t}$ for an outcome $y_t$, using the information available up to time $t-1$. The linear pool is the mixture
\begin{equation}
  F_t(y) \;=\; \sum_{n=1}^{N} w_{n,t}\, F_{n,t}(y), \qquad w_{n,t}\ge 0,\quad \sum_{n=1}^{N} w_{n,t}=1,
  \label{eq:linear_pool}
\end{equation}
where $w_{n,t}$ is the weight of the $n$th model and $f_t$ denotes the density of the pool. We call the $N$ component models the model set. The set is complete if it contains the model that generated the data, and incomplete otherwise \citep{Bernardo1994}. Much of the literature has focused on how to use Bayesian approaches to choose these weights. A prominent approach is Bayesian Model Averaging (BMA) \citep{Leamer1978,Raftery1997}, which weights the component models by their posterior probabilities. Although theoretically attractive, BMA is optimal only when the model set is complete. Since BMA weights accumulate evidence over the full sample, they adapt slowly to instability and structural change. These limitations have motivated combination methods that allow for an incomplete model set and let the weights vary over time. \citet{Hall2007} propose obtaining the combination weights by minimising the Kullback--Leibler divergence between the combined and the true density, which is equivalent to maximising the average log score and does not require the complete model set. \citet{Raftery2010} introduced dynamic model averaging (DMA), which discounts past evidence through a forgetting factor so that the weights track recent performance.  \citet{DelNegro2016} let the weights of a linear pool evolve as a latent stochastic process, and find that these dynamic pools react quickly to changes in the environment and outperform equal weights, BMA, optimal static pools and DMA in real time. \citet{Billio2013} give the combination weights a distributional state-space representation, with dynamics driven by the past performance of the predictive densities, and estimate them by sequential Monte Carlo. Their approach allows the model set to be incomplete and quantifies the uncertainty in the weights, which they find to be substantial when the predictors are similar. Bayesian predictive synthesis \citep{McAlinn2019} combines the forecasts through a dynamic latent factor model that adapts over time to their biases, miscalibration and inter-dependence. The Loss Discounting Framework (LDF) of \citet{Bernaciak2024} evaluates each model through a loss function and uses a multilevel discounting scheme that allows the weights to adapt flexibly. \citet{Diebold2023} maximise the log score subject to penalties that select a few forecasters and shrink their weights towards equality, and find that a simple average of a small, well-chosen group performs as well as or better than more complex weighting schemes.

The weights determine how much each model contributes to the combined forecast,
but not whether that forecast is reliable. Reliability is assessed through calibration and sharpness \citep{Gneiting2007, mitchell2011}. Calibration refers to the statistical consistency between the probabilistic forecasts and the observations. It is usually checked with the probability integral transform (PIT) of \citet{Dawid1984}
\begin{equation*}
    u_t = F_t(y_t),
\end{equation*}
where $F_t$ is the CDF in \eqref{eq:linear_pool} and $y_t$ is the observed value at time $t$. If $F_t$ is the true conditional distribution of $y_t$, the PITs are independent and uniformly distributed on $(0,1)$. Sharpness refers to the concentration or precision of probabilistic forecasts, which can be measured by the width of central prediction intervals. \citet{Gneiting2007} state that the goal of probabilistic forecasting is to maximise the sharpness subject to calibration. 

Proper scoring rules assess both properties together. A scoring rule assigns the reward $S(F_t, y_t)$ to a forecaster who issues the predictive distribution $F_t$ when $y_t$ is observed. It is proper if the expected reward is maximised by issuing the true distribution, so the forecaster has no incentive to misreport \citep{Gneiting2007}. Different proper scores emphasise different aspects of the forecast, so the choice of score should reflect how the forecast will be used. The logarithmic score $S(F_t,y_t)=\log f_t(y_t)$ evaluates the predictive density at the realised value. Maximising the expected log score is equivalent to minimising the Kullback--Leibler divergence from the true distribution \citep{Gneiting2007}, which makes it the natural criterion for likelihood-based combination \citep{Hall2007} and recalibration. The continuous ranked probability score (CRPS), $\mathrm{CRPS}(F_t,y_t)=\int_{-\infty}^{\infty}\{F_t(z)-\mathbbm{1}(y_t\le z)\}^2\,\mathrm{d}z$, compares the predictive CDF with the empirical CDF of the observation. It is less sensitive to tail behaviour than the log score
\citep{Gneiting2007}, and therefore provides a more robust check on overall accuracy. In many applications, however, the cost of a poor forecast is concentrated in a particular region, such as large exchange rate movements or periods of high inflation. Simply weighting the log score makes it improper. \citet{Diks2011} propose the conditional and censored likelihood scores, which focus on the region of interest while remaining proper, and \citet{dePunder2026} show how any strictly proper scoring rule can be localised in the same way.

Although linear pools perform well empirically, they are miscalibrated in general. Let $\mu_{n,t}$ and $\sigma^2_{n,t}$ denote the mean and variance of the $n$th component model. The linear pool then has mean and variance
\begin{equation}
    \bar{\mu}_t = \sum_{n=1}^{N} w_{n,t}\, \mu_{n,t}, \qquad
    \bar{\sigma}^2_t = \sum_{n=1}^{N} w_{n,t}\, \sigma^2_{n,t} + \sum_{n=1}^{N} w_{n,t}\,(\mu_{n,t}-\bar{\mu}_t)^2.
    \label{eq:poolmoments}
\end{equation}
The second term of $\bar{\sigma}^2_t$ is positive whenever the component means differ, so the variance of the pool exceeds the weighted average of the component variances. The gap grows as the component means become more diverse. Diversity among the components, which is usually the reason for combining, therefore also inflates the spread of the pooled distribution. This holds even in the ideal case where every component is calibrated. \citet{Hora2004} first showed that a linear combination of calibrated expert distributions is not itself calibrated. \citet{Ranjan2010} for binary event forecasts and \citet{Gneiting2013} for predictive distributions show that, under suitable
conditions, non-trivial linear combinations
of distinct calibrated forecasts are
over-dispersed. In practice, the individual components may also be misspecified \citep{Geweke2011, Gneiting2013}, which further distorts the location, spread or shape of the pooled distribution. These distortions may evolve over time as the combination weights change and as the data-generating process shifts. 

The miscalibration of linear pools has motivated methods that correct the pooled distribution after the weights have been chosen, typically by composing it with a time-invariant monotone transformation. \citet{Ranjan2010} and \citet{Gneiting2013} propose the beta-transformed linear pool (BLP), which composes the linear pool with a Beta CDF, and the simpler spread-adjusted linear pool. They show that both approaches correct the over-dispersion of the linear pool in forecasting S\&P~500 returns and daily temperature. \citet{Bassetti2018} replace the Beta CDF with a Dirichlet-process mixture of Beta distributions, which can represent asymmetric and multimodal corrections. In machine learning, \citet{Kuleshov2018} calibrate the predictive CDF of any regression model by composing it with an isotonic regression of the empirical frequencies on the predicted probabilities. The procedure is a static monotone map on the PIT scale, estimated once on a calibration set, and guarantees calibrated credible intervals given enough data.

These corrections have two limitations. The first concerns the family of maps. A two-parameter Beta CDF can correct errors in location and spread, but not distortions such as a multimodal PIT distribution. Mixtures and isotonic regression offer greater flexibility, but the former can be computationally intensive to update, while the latter lacks a smooth, finite-dimensional parametrisation suitable for score-driven updates. The second limitation is that a map estimated once corrects the miscalibration of the pool averaged over the sample. It is therefore inaccurate in every period in which the current miscalibration departs from that average.

Some recent methods let the correction change over time. \citet{van2024} update the parameters of the BLP using an online convex optimisation algorithm that minimises the CRPS. For non-stationary wind power data, their approach outperforms the best static combination chosen in hindsight. Adaptive conformal inference \citep{Gibbs2021,Zaffran2022} adjusts the nominal level of a prediction interval according to whether the previous interval covered the outcome. This keeps long-run coverage at its target under distribution shift, but it corrects intervals rather than the full predictive distribution.

In this paper, we recalibrate the pool by composing it with a strictly increasing map $\tau_\theta$ with parameter $\theta$ on the unit interval, so that the recalibrated forecast distribution is $\tau_{\theta}(F_t(y))$. We address the two limitations above in turn. For the family, besides the Beta CDF we use the monotone rational quadratic spline (RQS) of \citet{Durkan2019}. The RQS divides the unit interval into bins and fits a ratio of quadratics within each bin. Its flexibility is set by its number of bins, and it can represent multimodal distortions that a Beta CDF cannot. To let the map change over time, we update $\theta$ through a generalised autoregressive score (GAS) model \citep{creal2013,harvey2013}, which moves the parameters along the gradient of the latest log-likelihood. The recalibrated log score is $\log f_t(y_t) + \log\tau'_{\theta}(u_t)$, so the only term that depends on $\theta$ is a function of the PIT alone. Its gradient is therefore the GAS score, and each update moves the map in the direction that would have most improved the latest forecast.

A score-driven map, however, learns about a change in miscalibration only after the change has appeared in the observed PITs. Part of that change can be predicted earlier, because the distribution of the PIT depends on the current combined forecast, which is known before the outcome is observed. For example, when the weights shift after a regime change, the shape of the pool changes, and its miscalibration changes with it. Using observable features to predict and correct miscalibration is established in other settings. In regression, local recalibration methods estimate the PIT distribution conditionally on features of the input, so that the correction varies with the forecast being corrected \citep{Song2019,Kock2026}, and \citet{Klein2021} build calibration directly into deep distributional regression. Following this idea, we propose a convolution method that adds an inner map whose parameters are set by a neural network from features describing the shape of the current forecast. The GAS map is applied on top of the inner map and corrects the miscalibration that the shape of the forecast does not reveal. Combining score-driven filtering with neural networks has proved effective in related problems. For example, \citet{urettini2024} use a GAS filter to normalise non-stationary inputs to deep learning models.

We evaluate the methods in a simulation study covering several forms of miscalibration, and apply them to forecasts of nine G10 exchange rates and US inflation. In the simulations, the RQS map corrects multimodal miscalibration, which the Beta family cannot represent. When the miscalibration switches between regimes, the convolution method gives the largest log-score gains. In the applications, we recalibrate both the LDF pool and Bayesian predictive synthesis (BPS). Recalibration generally improves PIT uniformity, while gains in the log score and interval forecasts depend on the extent and nature of the original miscalibration.

The paper is organised as follows.  Section~\ref{sec:setup} sets out the recalibration framework. Section~\ref{sec:method} presents the map families and the time-varying recalibration methods. Section~\ref{sec:evaluation} describes the evaluation criteria. Section~\ref{sec:sim} reports the simulation study and Section~\ref{sec:empirical} the two applications. We discuss our approach and set out
directions for further research in Section~\ref{sec:conclusion}.

\section{Recalibration framework}
\label{sec:setup}
Let $G_t$, with density $g_t$, denote the true distribution of $y_t$ conditional on the information available at time $t-1$. If $F_t = G_t$, then $u_t \stackrel{\scriptsize \text{iid}}{\sim} \mathrm{Uniform}(0,1)$ \citep{Dawid1984, Diebold1998}. Otherwise, the PITs follow a distribution with CDF and density
\begin{equation}
  \Pi_t(u) = G_t\bigl(F_t^{-1}(u)\bigr),
  \qquad
  \pi_t(u) = \frac{g_t\bigl(F_t^{-1}(u)\bigr)}{f_t\bigl(F_t^{-1}(u)\bigr)},
  \qquad u \in (0,1).
  \label{eq:pitdist}
\end{equation}
The shape of $\pi_t$ reveals the form of the miscalibration. A predictive distribution that is too wide gives a hump-shaped $\pi_t$, one that is too narrow a U-shaped $\pi_t$, a biased one a monotone tilt, and one with several modes gives a multimodal $\pi_t$. Since both the weights and the data-generating process (DGP) change over time, $\pi_t$ in general changes as well.

We formulate recalibration as a transformation of the forecast distribution. Let $\tau_{\theta_t}: (0,1) \to (0,1)$ be a strictly increasing, differentiable map with $\tau_{\theta_t}(0) = 0$ and $\tau_{\theta_t}(1) = 1$, whose parameter $\theta_t$ is chosen using information available at $t-1$. The recalibrated predictive forecast distribution has CDF and density
\begin{equation*}
  \tilde F_t(y) \;=\; \tau_{\theta_t}\big(F_t(y)\big), \qquad \tilde f_t(y) \;=\; \tau_{\theta_t}'\big(F_t(y)\big)\, f_t(y),
  \label{eq:recal}
\end{equation*}
and the PIT of the recalibrated forecast is $v_t=\tilde F_t(y_t)=\tau_{\theta_t}(u_t)$. The derivative $\tau'_{\theta_t}$ is a probability density on $[0,1]$, so choosing a map is equivalent to choosing a density for the PITs, and the shape of $\tau_{\theta_t}$ describes the miscalibration it corrects. Taking logarithms of $\tilde f_t(y)$ at the realised value $y_t$ gives
\begin{equation}
  \log \tilde f_t(y_t) = \log f_t(y_t) + \ell_t,
  \qquad
  \ell_t = \log \tau'_{\theta_t}(u_t).
  \label{eq:loglik}
\end{equation}
The term $\ell_t$ is the log-score gain from recalibration. It depends on the outcome only through the PIT. It is positive when the map raises the predictive density at the outcome and negative when it lowers it, and it is zero for the identity map. 

The expected gain has a direct interpretation in terms of the Kullback--Leibler divergence between the true and the forecast distributions. By \eqref{eq:pitdist}, $\log \pi_t(u_t) = \log g_t(y_t) - \log f_t(y_t)$, so the log density of the PIT at the realised PIT equals the log-score gap between the true density and the forecast. \citet{berkowitz2001} shows that this identity holds for the log density of the inverse-normal transform of the PIT, and \citet{Bao2007} and \citet{Hall2007} use it to relate the Kullback--Leibler divergence of a density forecast from the true density to the divergence of its PIT. Since the divergence is invariant to invertible transformations, it equals the divergence of the PIT distribution from uniformity,
\begin{equation*}
  \mathrm{KL}\bigl(G_t \,\|\, F_t\bigr) = \mathbb{E}_{t-1}\bigl[\log g_t(y_t) - \log f_t(y_t)\bigr] = \mathbb{E}_{t-1}\bigl[\log \pi_t(u_t)\bigr] = \mathrm{KL}\bigl(\Pi_t \,\|\, \mathrm{U}\bigr),
  \label{eq:klic}
\end{equation*}
where $\mathrm{U}$ denotes the Uniform$(0,1)$ distribution. Following the same argument for the recalibrated forecast, we have
\begin{equation}
  \mathrm{KL}\bigl(G_t \,\|\, \tilde F_t\bigr) = \mathbb{E}_{t-1}\Bigl[\log \frac{\pi_t(u_t)}{\tau'_{\theta_t}(u_t)}\Bigr] = \mathrm{KL}\bigl(\Pi_t \,\|\, \tau_{\theta_t}\bigr),
  \label{eq:kl}
\end{equation}
the divergence of the PIT distribution from the distribution with CDF $\tau_{\theta_t}$. Subtracting the two divergences gives the expected log-score gain from recalibration,
\begin{equation}
  \mathbb{E}_{t-1}[\ell_t] = \mathrm{KL}\bigl(G_t \,\|\, F_t\bigr) - \mathrm{KL}\bigl(G_t \,\|\, \tilde F_t\bigr) = \mathrm{KL}\bigl(\Pi_t \,\|\, \mathrm{U}\bigr) - \mathrm{KL}\bigl(\Pi_t \,\|\, \tau_{\theta_t}\bigr),
  \label{eq:expgain}
\end{equation}
which is derived in~\ref{app:expectedlog}. The divergence in~\eqref{eq:kl} is non-negative and is zero if and only if $\tau_{\theta_t}' = \pi_t$, 
\begin{equation}
  \mathbb{E}_{t-1}[\ell_t] \;\leq\; \mathrm{KL}\bigl(\Pi_t \,\|\, \mathrm{U}\bigr),
  \qquad \text{with equality if and only if } \tau_{\theta_t} = \Pi_t.
  \label{eq:klbound}
\end{equation}
The optimal map is therefore the conditional CDF of the PIT itself, and the largest attainable gain is the divergence of the PIT distribution from uniformity, which measures how miscalibrated the pool is. A nearly calibrated pool leaves little to gain, whatever the map. A map that cannot reproduce $\pi_t$ leaves $\mathrm{KL}(\Pi_t \,\|\, \tau_{\theta_t})$ positive, and if this exceeds $\mathrm{KL}(\Pi_t \,\|\, \mathrm{U})$, the recalibrated forecast scores worse than the original one. In practice $\Pi_t$ is unknown and has to be learned from past PITs. Since the first term in \eqref{eq:expgain} does not depend on the map, the best map at time $t$ maximises the expected gain given the information at $t-1$,
\begin{equation}
\theta_t^{*} = \arg\max_{\theta}\, \mathbb{E}_{t-1}\!\left[\log \tau'_{\theta}(u_t)\right].
\label{eq:target}
\end{equation}
Maximising the expected gain is the same as minimising $\mathrm{KL}(\Pi_t \,\|\, \tau_{\theta})$, so $\theta_t^{*}$ gives the map whose implied PIT distribution is closest to $\Pi_t$.

\section{Methodology}
\label{sec:method}
\subsection{Map families}
\label{sec:maps}
\subsubsection{Beta map}
\label{sec:beta}
Following the BLP of \citet{Gneiting2013}, the first family is the Beta CDF. The recalibrated PIT is
\begin{equation*}
  v_t = \tau_{\theta_t}(u_t) = I_{u_t}(a_t, b_t) = \int_0^{u_t} \beta(x; a_t, b_t)\, dx,
  \label{eq:beta_map}
\end{equation*}
where $I_{u}(a, b)$ is the regularised incomplete beta function and
\begin{equation*}
  \beta(u; a, b) = \frac{u^{a-1}(1-u)^{b-1}}{B(a, b)},
  \quad
  B(a,b) = \frac{\Gamma(a)\Gamma(b)}{\Gamma(a+b)},
  \quad u \in (0,1), \quad a,b >0.
  \label{eq:beta_pdf}
\end{equation*}
The derivative of the map is the beta density $\tau_{\theta_t}'(u) = \beta(u; a_t, b_t)$, so substituting it into \eqref{eq:loglik} and taking logarithms gives the log-score gain
\begin{equation}
  \ell_t = \log \beta(u_t; a_t, b_t) = (a_t - 1)\log u_t + (b_t - 1)\log(1 - u_t) - \log B(a_t, b_t).
  \label{eq:beta_ll}
\end{equation}
We update an unconstrained parameter $\theta_t = (\theta_{a,t}, \theta_{b,t})' \in \mathbb{R}^2$ and map it to the beta parameters by
\begin{equation}
  a_t = c + \mathrm{softplus}(\theta_{a,t} + s),
  \qquad
  b_t = c + \mathrm{softplus}(\theta_{b,t} + s),
  \qquad
  s = \log\bigl(e^{1 - c} - 1\bigr),
  \label{eq:beta_link}
\end{equation}
where $\mathrm{softplus}(x) = \log(1 + e^{x})$ and $c = 0.05$. The floor $c$ keeps both beta parameters bounded away from zero to prevent numerical instability. The shift $s$ satisfies $\mathrm{softplus}(s) = 1 - c$, so that $\theta_t = 0$ corresponds exactly to $a_t = b_t = 1$, the identity map.

The score-driven update requires the derivative of the log-score gain with respect to the unconstrained parameter, that is,
\begin{equation}
\nabla_t = \frac{\partial \ell_t}{\partial \theta_t} =
\begin{pmatrix}
\bigl[\log u_t - \psi(a_t) + \psi(a_t + b_t)\bigr]\Lambda(\theta_{a,t} + s) \\[2pt]
\bigl[\log(1 - u_t) - \psi(b_t) + \psi(a_t + b_t)\bigr]\Lambda(\theta_{b,t} + s)
\end{pmatrix},
\label{eq:beta_score}
\end{equation}
where $\psi$ is the digamma function and $\Lambda(x) = 1/(1+e^{-x})$ is the logistic function. 
The full derivation is given in~\ref{app:beta_score}.

\subsubsection{Rational quadratic spline map}
The second family is the monotone rational quadratic spline (RQS) of \citet{Durkan2019}. The spline map divides the unit interval into $K$ bins. Within each bin the map is a ratio of two quadratic polynomials whose value and derivative are fixed at the bin edges (the knots), which makes the map strictly increasing and continuously differentiable on $(0,1)$. The map, its derivative and its inverse are all available in closed form. The spline is determined by $K$ bin widths, $K$ bin heights and positive derivatives at the $K+1$ knots, including the two end points. These are obtained from an unconstrained parameter
\begin{equation*}
\theta_t = (\theta^{u}_{1,t}, \dots, \theta^{u}_{K,t}, \theta^{v}_{1,t}, \dots, \theta^{v}_{K,t}, \theta^{d}_{0,t}, \dots, \theta^{d}_{K,t}) \in \mathbb{R}^{3K+1},
\end{equation*}
through the transformations
\begin{equation}
w_{k,t} = \mathrm{softmax}(\theta^u_{t})_k, \quad h_{k,t} = \mathrm{softmax}(\theta^v_{t})_k, \quad\delta^{(k)}_t = m+\mathrm{softplus}_{\varrho}(\theta_{k,t}^d),   
\label{eq:rqs_link}
\end{equation}
where $\mathrm{softmax}(x)_k = e^{x_k}/\sum_{j=1}^{K} e^{x_j}$, $\mathrm{softplus}_{\varrho}(x) = \varrho^{-1}\log(1 + e^{\varrho x})$, $\varrho = \log 2/(1 - m)$ and $m = 10^{-3}$. The widths $w_{k,t}$ and heights $h_{k,t}$ are positive and each sums to one, and the floor $m$ prevents any derivative from collapsing to zero. The knot positions are then obtained by cumulative summation:
\begin{equation*}
u_{t}^{(k)} = \sum_{j=1}^k w_{j,t}, \quad v_{t}^{(k)} = \sum_{j=1}^k h_{j,t},
\end{equation*}
so that $u^{(0)}_{t} = v^{(0)}_{t} = 0$ and $u^{(K)}_{t} = v^{(K)}_{t} = 1$, and the map sends each input bin $[u^{(k)}_t, u^{(k+1)}_t]$ onto the output bin $[v^{(k)}_t, v^{(k+1)}_t]$. At $\theta_t = 0$, all widths and heights equal $1/K$, and the scale $\varrho$ makes every knot derivative equal to one, so $\theta_t = 0$ again gives the identity map. We index the bins by $k=0,\dots,K-1$. For $u_t$ in the $k$-th bin, let
\begin{equation*}
\xi_t = \frac{u_t - u^{(k)}_t}{u^{(k+1)}_t - u^{(k)}_t}
 \quad \text{and} \quad s^{(k)}_t = \frac{v^{(k+1)}_t - v^{(k)}_t}{u^{(k+1)}_t - u^{(k)}_t}    
\end{equation*}
denote the relative position within the bin and the bin’s average slope. The recalibrated PIT is
\begin{equation*}
    v_t = v^{(k)}_t +\frac{(v^{(k+1)}_{t} - v^{(k)}_{t})\bigl[s^{(k)}_t \xi_t^2 + \delta^{(k)}_t\xi_t(1-\xi_t)\bigr]}{s^{(k)}_t  + \bigl[\delta^{(k+1)}_t + \delta^{(k)}_t - 2s^{(k)}_t\bigr]\xi_t(1-\xi_t)}.
    \label{eq:rqs}
\end{equation*}
Differentiating with respect to $u_t$ and taking logarithms gives the log-score gain
\begin{equation}
    \ell_t =\log\frac{\big(s^{(k)}_t\big)^2\Big[\delta^{(k+1)}_t\xi_t^2 + 2s^{(k)}_t\xi_t(1-\xi_t) + \delta^{(k)}_t(1-\xi_t)^2 \Big]}{\bigl(s^{(k)}_t + (\delta^{(k+1)}_t + \delta^{(k)}_t - 2s^{(k)}_t)\xi_t(1-\xi_t)\bigr)^2}.
    \label{eq:rqs_lll}
\end{equation}
The log-score gain \eqref{eq:rqs_lll} depends on $\theta_t$ through four quantities $\xi_t,\ s^{(k)}_t,\ \delta^{(k)}_t,\ \delta^{(k+1)}_t$. The score $\nabla_t = \partial \ell_t/\partial \theta_t$  therefore follows from the chain rule in three stages: the derivatives of $\ell_t$ with respect to these four quantities, their derivatives with respect to the widths, heights and knot derivatives, and the derivatives of those with respect to $\theta_t$ through the softmax and softplus links \eqref{eq:rqs_link}. All three stages are available in closed form\footnote{In practice we evaluate the score by automatic differentiation, which yields the same results.} and are given in~\ref{app:rqs_score}.

\subsection{Score-driven recalibration}
\label{sec:gas}
\subsubsection{A two-timescale score-driven filter}
The parameter of the map evolves by a score-driven recursion \citep{creal2013,harvey2013}. In its standard form, the GAS update is
\begin{equation}
  \theta_{t+1} = \omega + B\,\theta_t + A\, s_t,
  \qquad
  s_t = S_t \nabla_t,
  \label{eq:gas-standard}
\end{equation}
where $\nabla_t = \partial \ell_t / \partial \theta_t$ is the score of the log-score gain \eqref{eq:loglik} at the latest PIT, $S_t$ is a scaling matrix, often an inverse power of the Fisher information, and $\omega$, $A$ and $B$ set the level, the response to new information and the persistence of the parameter.

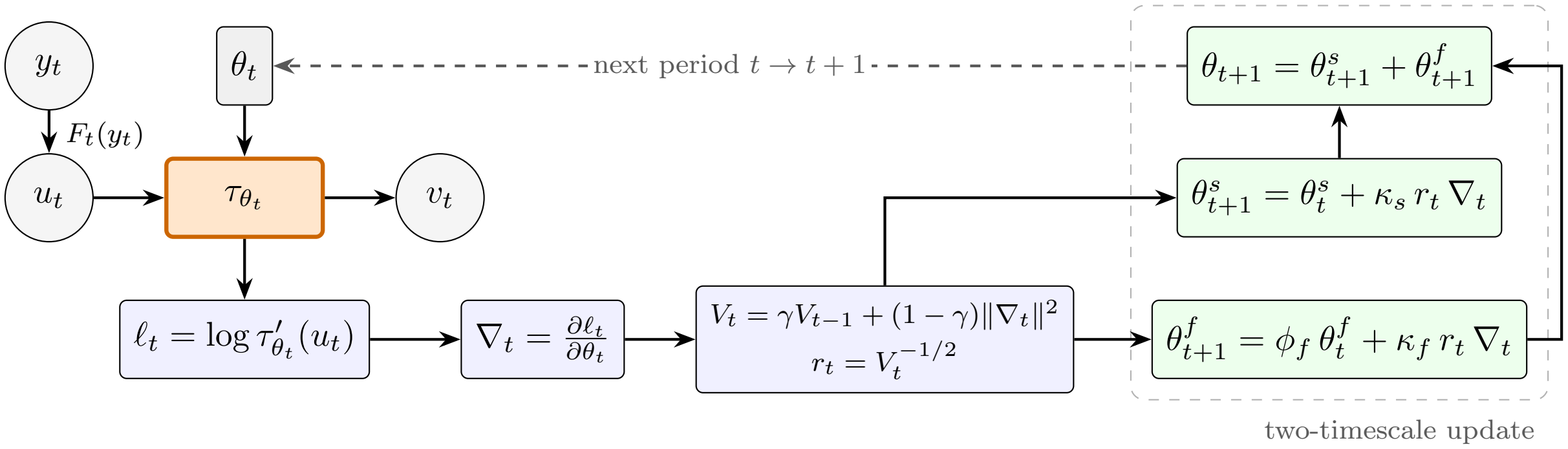
\begin{figure}[H]
\centering
\begin{tikzpicture}[
  font=\footnotesize,
  >={Stealth[length=2.2mm]},
  box/.style   ={draw, rounded corners=2pt, minimum height=8mm, inner sep=4pt, align=center},
  data/.style  ={box, circle, minimum size=9mm, inner sep=1pt, fill=gray!8},
  map/.style   ={box, very thick, draw=orange!80!black, fill=orange!20, minimum width=16mm},
  score/.style ={box, fill=blue!6},
  state/.style ={box, fill=green!7},
  param/.style ={box, fill=gray!12},
  grp/.style   ={draw=gray!60, dashed, rounded corners=4pt, inner sep=6pt},
  fl/.style    ={->, thick},
  flb/.style   ={->, thick, dashed, gray!70!black},
  lbl/.style   ={font=\scriptsize, fill=white, inner sep=1.5pt},
]
\node[param] (thetat) at (0,1.35)   {$\theta_t$};
\node[data] (y)       at (-2,1.35)     {$y_t$};
\node[data]  (u)      at (-2.0,0)   {$u_t$};
\node[map]   (tau)    at (0,0)      {$\tau_{\theta_t}$};
\node[data]  (v)      at (2.0,0)    {$v_t$};
\node[score] (ll) at (0,-1.45)   {$\ell_t=\log\tau'_{\theta_t}(u_t)$};
\node[score] (g)  at (3.05,-1.45) {$\nabla_t=\frac{\partial \ell_t}{\partial \theta_t}$};
\node[score] (r)  at (6.55,-1.45)
  {\scriptsize $V_t=\gamma V_{t-1}+(1-\gamma)\lVert\nabla_t\rVert^2$\\[2pt]\scriptsize $r_t=V_t^{-1/2}$};
\node[state] (fast)  at (11.2,-1.45) {$\theta^f_{t+1}= \phi_f\,\theta^f_t+ \kappa_f\,r_t\,\nabla_t$};
\node[state] (slow)  at (11.2,0)     {$\theta^s_{t+1}=\theta^s_t+\kappa_s\,r_t\,\nabla_t$};
\node[state] (theta) at (11.2,1.35)  {$\theta_{t+1}=\theta^s_{t+1}+\theta^f_{t+1}$};
\begin{scope}[on background layer]
  \node[grp, fit=(slow)(fast)(theta)] (upd) {};
\end{scope}
\node[font=\scriptsize, gray!70!black, anchor=north east]
  at ($(upd.south east)+(0,-0.05)$) {two-timescale update};

\draw[fl] (y) -- (u) node[lbl, pos=0.58,right=3pt]{$F_t(y_t)$};
\draw[fl] (u) -- (tau);
\draw[fl] (thetat) -- (tau);
\draw[fl] (tau) -- (v);
\draw[fl] (tau) -- (ll);
\draw[fl] (ll) -- (g);
\draw[fl] (g) -- (r);
\draw[fl] (r.north) |- (slow.west);
\draw[fl] (r.east) -- (fast.west);
\draw[fl] (slow) -- (theta);
\draw[fl] (fast.east) -- ++(0.35,0) |- (theta.east);
\draw[flb] (theta.west) -- (thetat.east)
  node[lbl, pos=0.5] {next period $t\to t+1$};
\end{tikzpicture}
\caption{Overview of the score-driven filter. Solid arrows: computation within period $t$; dashed arrows: updates carried to later periods.}
\label{fig:gas}
\end{figure}

We modify \eqref{eq:gas-standard} in two ways. First, since the Fisher information has no closed form for the spline map, we scale the score by a running root-mean-square normaliser for both families,
\begin{equation*}
  V_t = \gamma\,V_{t-1} + (1-\gamma)\,\lVert \nabla_t \rVert^2, 
  \qquad
  r_t = V_t^{-\frac{1}{2}}, \qquad V_0 = 10^{-4},
\end{equation*}
where $\gamma \in (0,1)$ is a decay rate. The size of a step then does not depend on the scale of the score, which differs between the two families and across the unit interval. Second, we write the parameter as the sum of a slow and a fast component,
\begin{equation}
  \theta_t = \theta^{s}_t + \theta^{f}_t,
  \qquad
  \theta^{s}_{t+1} = \theta^{s}_t + \kappa_s\, r_t \nabla_t,
  \qquad
  \theta^{f}_{t+1} = \phi_f\, \theta^{f}_t + \kappa_f\, r_t \nabla_t,
  \quad |\phi_f| < 1.
  \label{eq:twotimescale}
\end{equation}
The slow component $\theta^{s}_t$ has a random-walk-type recursion that accumulates persistent miscalibration, and the fast
component $\theta^{f}_t$ follows a mean-reverting score-driven recursion with AR(1)-type persistence that responds to transient departures and decays back to zero. A single filter can therefore track both a permanent and a temporary shift in calibration without forcing them to share one adaptation rate. Since $\theta_t$ is a deterministic function of $u_1, \dots, u_{t-1}$, $\ell_t$ is a one-step-ahead out-of-sample log-score gain for every $t$. Figure~\ref{fig:gas} illustrates the filter.

\subsubsection{Dynamic model averaging of filters}
\label{sec:dma}
The adaptation rates $\gamma$, $\kappa_s$, $\kappa_f$, $\phi_f$ control how quickly the map responds to new PITs. The right speed of adaptation is unknown and may itself change over time. Rather than commit to a single set of rates, we therefore run $E = 5$ filters in parallel, each with fixed rates, and combine them by dynamic model averaging (DMA). The rates, listed in Table~\ref{tab:experts}, range from an identity filter that never moves to a rapidly adapting filter. The DMA weights learn from past performance which speed of adaptation suits the current miscalibration, and they can shift between filters as it changes.

Each filter $e$ has its own parameter $\theta_{e,t}$, updated by \eqref{eq:twotimescale}, and hence its own map $\tau_{\theta_{e,t}}$. The map applied at time $t$ is the mixture of the five maps, with weights $\omega_{e,t\mid t-1}$ that depend only on PITs up to $t-1$. The recalibrated PIT and its log-score gain are
\begin{equation*}
  v_t \;=\; \sum_{e=1}^{E} \omega_{e,t\mid t-1}\, \tau_{\theta_{e,t}}(u_t),
  \qquad
  \ell_t \;=\; \log \sum_{e=1}^{E} \omega_{e,t\mid t-1}\,
               \tau_{\theta_{e,t}}'(u_t).
  \label{eq:mixture}
\end{equation*}
We write $\tau_{\theta^{\mathrm{GAS}}_t}$ for this mixture map, so that $v_t = \tau_{\theta^{\mathrm{GAS}}_t}(u_t)$ and $\ell_t = \log \tau'_{\theta^{\mathrm{GAS}}_t}(u_t)$, where $\theta^{\mathrm{GAS}}_t$ collects the filter parameters and weights.

Once $y_t$ is observed, the weights are updated from each filter's own gain $\ell_{e,t} = \log \tau_{\theta_{e,t}}'(u_t)$ by DMA \citep{Raftery2010},
\begin{equation}
    \omega_{e,t\mid t} = \frac{\omega_{e,t\mid t-1} \exp(\ell_{e,t})}{\sum_{j=1}^{E}\omega_{j,t\mid t-1} \exp(\ell_{j,t})},
    \qquad
    \omega_{e,t+1\mid t} = \frac{\omega_{e,t\mid t}^{\,\alpha^{\mathrm{DMA}}}+c_{\omega}}{\sum_{j=1}^{E}\bigl(\omega_{j,t\mid t}^{\,\alpha^{\mathrm{DMA}}}+c_{\omega}\bigr)},
    \label{eq:dma}
\end{equation}
starting from $\omega_{e,1\mid 0} = 1/E$. The first step is a Bayesian update of the weights given the latest PIT. The second discounts past evidence through the forgetting factor $\alpha^{\mathrm{DMA}} \in (0,1]$, which plays the same role as the discount factor in the LDF, and adds a small constant $c_{\omega} > 0$ to avoid numerical instability.

\begin{table}[H]
  \centering
  \begin{tabular}{lcccc}
    \toprule
    Filter & $\kappa_s$ & $\kappa_f$ & $\phi_f$ & $\gamma$ \\
    \midrule
    Identity & 0.00 & 0.00 & 0.00 & 0.93 \\
    Glacial  & 0.005 & 0.00 & 0.90 & 0.93 \\
    Slow     & 0.03 & 0.00 & 0.90 & 0.93 \\
    Medium   & 0.05 & 0.15 & 0.85 & 0.93 \\
    Rapid    & 0.08 & 0.40 & 0.60 & 0.93 \\
    \bottomrule
  \end{tabular}
  \caption{Adaptation rates of the filters.}
  \label{tab:experts}
\end{table}

\subsection{Convolution}
\label{sec:convolution}
As discussed in Section~\ref{sec:intro}, the shape of the current forecast can reveal how it is miscalibrated, because the pool changes shape as its weights shift. The convolution method exploits this by composing two maps from the same family: an inner map whose parameter is set from the shape of $F_t$, and the mixture map of Section~\ref{sec:gas}. The recalibrated forecast is
\begin{equation}
  \tilde F_t(y) = \tau_{\theta^{\mathrm{GAS}}_t}\!\Bigl(\tau_{\theta^{\mathrm{NN}}_t}\bigl(F_t(y)\bigl)\Bigr),
  \qquad
  \theta^{\mathrm{NN}}_t = h(\mathbf{z}_t; W),
  \label{eq:composition}
\end{equation}
where $\mathbf{z}_t$ is a vector of features describing the shape of $F_t$, and $h$ is a neural network with weights $W$. Writing $\tilde u_t = \tau_{\theta^{\mathrm{NN}}_t}(u_t)$ for the output of the inner map, the recalibrated PIT and its log-score gain are
\begin{equation}
  v_t = \tau_{\theta^{\mathrm{GAS}}_t}(\tilde u_t),
  \qquad
  \ell_t = \log \tau'_{\theta^{\mathrm{NN}}_t}(u_t) + \log \tau'_{\theta^{\mathrm{GAS}}_t}(\tilde u_t).
  \label{eq:composition-gain}
\end{equation}
The inner map corrects the part of the miscalibration that can be predicted from the shape of the forecast, and the filters correct what remains (Figure~\ref{fig:conv}). The rest of this section describes the features $\mathbf{z}_t$ and how $W$ is estimated.

\subsection*{Predictive shape features}
The features used in the neural network should capture the aspects of the forecast distribution's shape that determine how it is miscalibrated: its location, its dispersion in the centre and in the tails, its asymmetry, and the weight of its tails. We compute them from $F_t$ on the scale of the outcome, before $y_t$ is observed. On the PIT scale every forecast is mapped to the uniform distribution by its own CDF, so all forecasts would have the same shape. We measure each aspect with quantiles $q_{t,p} = F_t^{-1}(p)$, which exist for any predictive distribution and are insensitive to small amounts of probability far in the tails. We use $p \in \{0.05,\allowbreak 0.125,\allowbreak 0.25,\allowbreak 0.375,\allowbreak 0.5,\allowbreak 0.625,\allowbreak 0.75,\allowbreak 0.875,\allowbreak 0.95\}$, evaluating $F_t$ on a fixed grid $x_1 < \dots < x_J$ and obtaining the quantiles by linear interpolation. 

Table~\ref{tab:quantile-features} defines the eight features, using the interquartile range $\mathrm{IQR}_t = q_{t,0.75} - q_{t,0.25}$,  the left and right half-spreads of the 90\% interval, $D^{L}_t = q_{t,0.5} - q_{t,0.05}$ and $D^{R}_t = q_{t,0.95} - q_{t,0.5}$. $X_1$ measures location. $X_2$ and $X_3$ measure dispersion over the central half and the central 90\% of the distribution, which allows the network to distinguish a change in the centre of the forecast distribution from a change in its tails. $X_4$ and $X_5$ are the quantile-based skewness and kurtosis of \citet{bowley1901} and \citet{moors1988}. $X_6$ and $X_7$ measure the length of each tail relative to the interquartile range, and $X_8$ measures the asymmetry between the two tails.

\begin{table}[H]
  \centering
  \footnotesize
  \begin{tabular}{@{}l c l@{}}
    \toprule
    Variable & Definition & Description \\
    \midrule
    $X_{1,t}$ & $(q_{t,0.50} - x_1)/(x_J - x_1)$ & median location \\[9pt]
    $X_{2,t}$ & $\log \mathrm{IQR}_t$ & log interquartile range \\[9pt]
    $X_{3,t}$ & $\log (q_{t,0.95} - q_{t,0.05})$ & log $90\%$ spread \\[9pt]
    $X_{4,t}$ & $(q_{t,0.75} + q_{t,0.25} - 2\,q_{t,0.50})/{\mathrm{IQR}_t}$ & Bowley skewness \\[9pt]
    $X_{5,t}$ & $[(q_{t,0.875} - q_{t,0.625}) + (q_{t,0.375} - q_{t,0.125})]/{\mathrm{IQR}_t}$ & Moors kurtosis \\[9pt]
    $X_{6,t}$ & $D^{L}_t/{\mathrm{IQR}_t}$ & scaled left half-spread \\[11pt]
    $X_{7,t}$ & $D^{R}_t/{\mathrm{IQR}_t}$ & scaled right half-spread \\[11pt]
    $X_{8,t}$ & $(D^{R}_t - D^{L}_t)/({D^{R}_t+ D^{L}_t})$ & tail asymmetry \\
    \bottomrule
  \end{tabular}
  \caption{Quantile-based features of the predictive distribution $F_t$.}
  \label{tab:quantile-features}
\end{table}

\subsection*{Causal rank normalisation}
The features differ in units and scale, both from one another and between the variables being forecast. We therefore standardise each feature against its own past, so that the same network architecture and training settings can be used for any forecast target,
\begin{equation*}
\begin{aligned}
z_{j,t} &=
\begin{cases}
0, & t \le t_0,\\[4pt]
\Phi^{-1}\!\left(
\dfrac{(t-1)\hat r_{j,t}+0.5}{t}
\right), & t>t_0,
\end{cases}\\[8pt]
\hat r_{j,t} &=
\frac{1}{t-1}\sum_{s=1}^{t-1}
\left[
\mathbbm{1}\{X_{j,s}<X_{j,t}\}
+\frac12\mathbbm{1}\{X_{j,s}=X_{j,t}\}
\right], \quad t>t_0.
\end{aligned}
\label{eq:causal_rank}
\end{equation*}
We set $t_0=20$ and initialise $z_{j,t}=0$ for $t\leq t_0$, corresponding to the median of the standard normal distribution, since the available history is too short to obtain reliable empirical ranks. For $t>t_0$, the features are transformed using their ranks relative to previous observations. The rank $\hat{r}_{j,t}$ is the fraction of past values of feature $j$ below its current value, with equal values counted as half, so it records where the current forecast stands relative to earlier forecasts. The adjustment keeps the argument of $\Phi^{-1}$ in the open interval $(0,1)$, so that $z_{j,t}$ is always finite and expressed on a standard normal quantile scale. A value of $z_{j,t}$ near zero means that feature $j$ is at its historical median, while large positive or negative values mean that the current forecast is unusually high or low on that feature relative to its own history. For example, $z_{3,t} = 1.28$ means that the 90\% spread of $F_t$ exceeds about 90\% of its past values. The network therefore responds to how unusual the current shape of the forecast is, not to its absolute level.

\subsection*{Neural network estimation}

The network $h$ in \eqref{eq:composition} maps
$\mathbf{z}_t=(z_{1,t},\dots,z_{8,t})'$ to
$\theta^{\mathrm{NN}}_t\in\mathbb{R}^{\dim_\theta}$
through one hidden layer of $H$ ReLU units, where
$\dim_\theta=2$ for the Beta map and $\dim_\theta=3K+1$
for the spline. The network weights $W$ are estimated
by numerical optimisation, while the GAS filter parameters and DMA weights evolve according to
\eqref{eq:twotimescale} and \eqref{eq:dma}.

At each period $t$, the network determines the inner
map from the features of $F_t$. Once $y_t$ is observed,
the inner map transforms $u_t$ into
$\tilde u_t=\tau_{\theta^{\mathrm{NN}}_t}(u_t)$,
which is subsequently used to update the GAS filters
and their DMA weights for the next period.
The resulting recalibrated forecast and log-score gain
are given by \eqref{eq:composition} and
\eqref{eq:composition-gain}.

\begin{figure}[H]
\centering
\resizebox{\linewidth}{!}{%
\begin{tikzpicture}[
  font=\footnotesize,
  >={Stealth[length=2.2mm]},
  box/.style   ={draw, rounded corners=2pt, minimum height=8.5mm, inner sep=4pt, align=center},
  data/.style  ={box, circle, minimum size=9mm, inner sep=1pt, fill=gray!8},
  map/.style   ={box, very thick, draw=orange!80!black, fill=orange!20, minimum width=15mm},
  feat/.style  ={box, fill=blue!6, minimum width=17mm},
  state/.style ={box, fill=green!7},
  param/.style ={box, fill=gray!12},
  score/.style ={box, fill=red!6},
  grp/.style   ={draw=gray!55, dashed, rounded corners=4pt, inner sep=6pt},
  grplbl/.style={font=\scriptsize\itshape, gray!70!black},
  fl/.style    ={->, thick},
  flb/.style   ={->, thick, dashed, gray!70!black},
  lbl/.style   ={font=\scriptsize, fill=white, inner sep=1.5pt},
]
\node[data]  (F)    at (0,1.75)    {$F_t$};
\node[feat]  (X)    at (2,1.75)  {$X_t$\\[-1pt]{\scriptsize shape}\\[-1pt]{\scriptsize features}};
\node[feat]  (z)    at (4.5,1.75)  {$\bm z_t$\\[-1pt]{\scriptsize causal}\\[-1pt]{\scriptsize ranks}};
\node[feat]  (h)    at (7,1.75)  {$h(\bm z_t;\hat W_t)$\\[-1.5pt]{\scriptsize network}};
\node[state] (bank) at (12,1.75) {$E$ score-driven filters};

\node[data] (y)    at (-1.5,0)  {$y_t$};
\node[data] (u)    at (0,0)     {$u_t$};
\node[map]  (tauN) at (7,0)   {$\tau_{\theta_t^\mathrm{NN}}$\\[-1pt]{\scriptsize inner map}};
\node[data] (ut)   at (9.4,0)   {$\tilde u_t$};
\node[map]  (tauG) at (12,0)    {$\tau_{\theta_t^\mathrm{GAS}}$\\[-1pt]{\scriptsize outer map}};
\node[data] (v)    at (14.7,0)  {$v_t$};

\node[score] (ellN) at (7,-1.7) {$\ell^{\mathrm{NN}}_t = \log\tau_{\theta_t^\mathrm{NN}}'(u_t)$};
\node[score] (ellG) at (12,-1.7)  {$\ell^{\mathrm{GAS}}_t = \log \tau_{\theta_t^\mathrm{GAS}}'(\tilde u_t)$};
\node[score] (ell)  at (9.6,-3) {$\ell_t = \ell^{\mathrm{NN}}_t + \ell^{\mathrm{GAS}}_t$};

\node[param] (train) at (7,3.6)
  {$\hat W_t$};

\begin{scope}[on background layer]
  \node[grp, fit=(X)(z)(h)(tauN)] (gin)  {};
  \node[grp, fit=(bank)(tauG)]    (gout) {};
\end{scope}
\node[grplbl, anchor=north west] at (gin.south west)  {inner map: covariate-driven};
\node[grplbl, anchor=south east] at (gout.north east) {outer map: score-driven};

\draw[fl] (F) -- (X);
\draw[fl] (X) -- (z);
\draw[fl] (z) -- (h);
\draw[fl] (F) -- (u);
\draw[fl] (y) -- (u);
\draw[fl] (h) -- (tauN);
\draw[fl] (u) -- (tauN);
\draw[fl] (tauN) -- (ut);
\draw[fl] (bank) -- (tauG)
  node[lbl, midway, right=2pt] {$\omega_{e,t\mid t-1},\,\theta_{e,t}$};
\draw[fl] (ut) -- (tauG);
\draw[fl] (tauG) -- (v);
\draw[fl] (tauN) -- (ellN);
\draw[fl] (tauG) -- (ellG);
\draw[fl] (ellN) |- (ell);
\draw[fl] (ellG) |- (ell);

\draw[flb] (train) -- (h);
\draw[flb] (ut.north) |- (bank.west)
  node[lbl, align=center, pos=0.2, left=1pt] {update \\[-1pt]for $t{+}1$};
\end{tikzpicture}}
\caption{Overview of the two-stage recalibration procedure. Shape features of $F_t$ set the inner map, whose output $\tilde u_t$ is recalibrated by the score-driven filters of Figure~\ref{fig:gas}. Solid arrows: computation within period $t$; dashed arrows: updates carried to later periods.}
\label{fig:conv}
\end{figure}
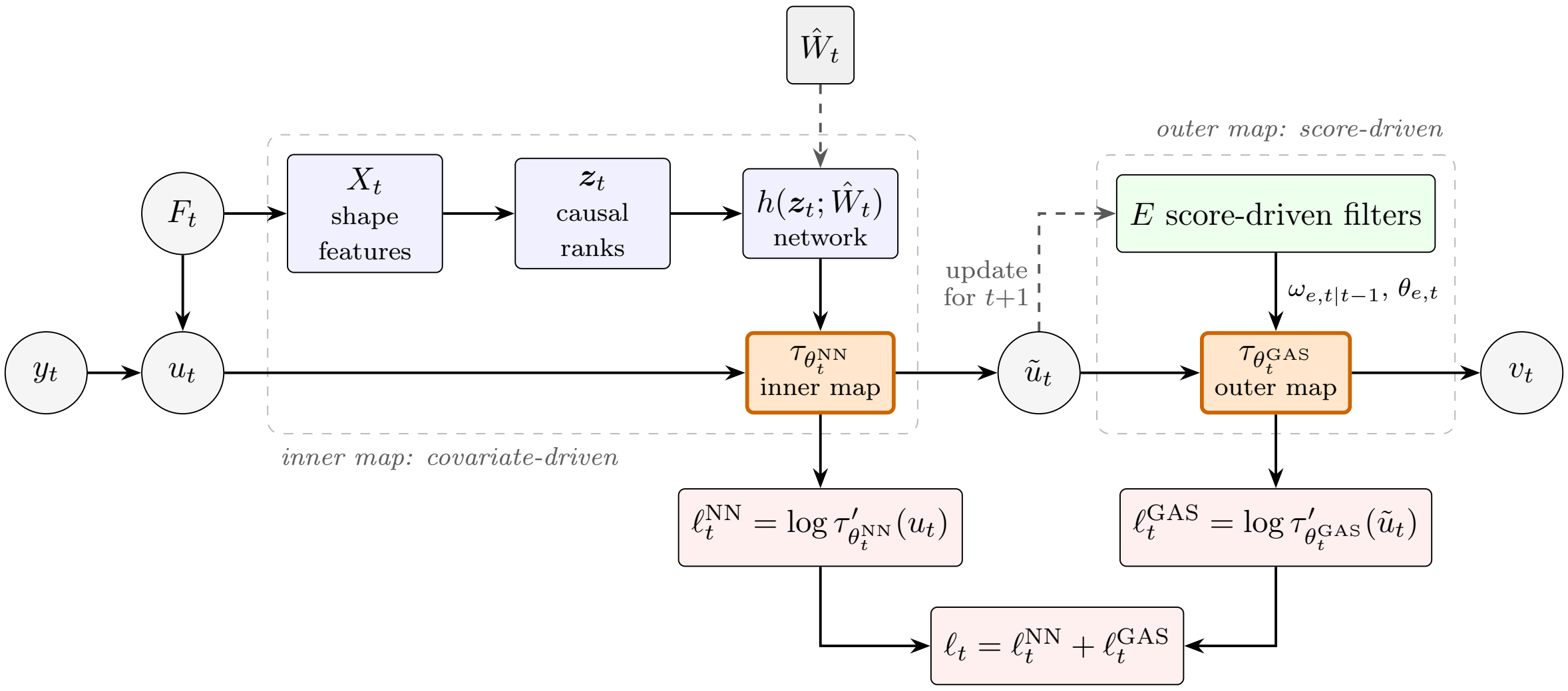

The network is re-estimated every $R$ periods on an
expanding window. At each refit, the historical GAS
filter parameters and DMA weights are held fixed, and the network
is fitted by maximising the conditional composite
log-score gain,
\begin{equation}
\hat W_t=\arg\max_W
\sum_{s=t_{\mathrm w}}^{t_{\mathrm{tr}}-1}
\Bigg[
\log\tau'_{h(\mathbf z_s;W)}(u_s)
+\log\sum_{e=1}^{E}\omega_{e,s\mid s-1}
\tau'_{\theta_{e,s}}
\left(\tau_{h(\mathbf z_s;W)}(u_s)\right)
\Bigg],
\label{eq:trainobj}
\end{equation}
where $t_{\mathrm w}$ denotes the start of the
training window and $t_{\mathrm{tr}}$ the start
of the validation window. The network is trained
using Adam with weight decay, with the earliest
75\% of observations used for optimisation and
the remaining 25\% for chronological validation.
Early stopping is applied after 40 epochs without
improvement in the validation gain. At each refit,
the network is reinitialised near the identity,
while the GAS filters and DMA weights continue
without resetting. Figure~\ref{fig:conv} summarises
the procedure.

\section{Evaluation}
\label{sec:evaluation}

\subsection{Calibration}
\label{sec:eval-calibration}
We assess calibration by testing whether the PITs are uniform and independent, using several tests that are sensitive to different departures. \citet{Gneiting2007} define probabilistic calibration of a sequence of forecasts $F_t$, relative to the true DGP $G_t$, as the condition
\begin{equation*}
    \frac{1}{T} \sum_{t=1}^T G_t \big(F_t^{-1}(p) \big) \rightarrow p \quad \text{for all} \quad p \in (0,1),
\end{equation*}
which is equivalent to uniformity of the PITs. Uniformity says nothing about dependence over time. One-step-ahead forecasts that use all available information also have independent PITs \citep{Dawid1984, Diebold1998}, and \citet{mitchell2011} call uniformity and independence together \emph{complete calibration}. We test uniformity of the PITs with four tests. The Kolmogorov--Smirnov (KS) test measures the largest distance between the empirical CDF of the PITs and the uniform CDF. The Anderson--Darling (AD) test weights departures near 0 and 1 more heavily, so it is more sensitive to miscalibration in the tails. Pearson's $\chi^2$ test compares bin counts with their expected values under uniformity, $B$ equal-width bins in all, and so formalises the PIT histograms. The Doornik--Hansen (DH) test \citep{doornik2008} checks the normality of $\zeta_t = \Phi^{-1}(v_t)$ through its skewness and kurtosis. It detects distortions in the shape of the predictive distribution that can leave the PIT histogram roughly flat. We test independence with the Ljung--Box (LB) test on the PIT autocorrelations at lags one to four. The test of \citet{berkowitz2001} fits a Gaussian AR(1) to $\zeta_t$ and tests jointly for zero mean, unit variance and no first-order autocorrelation, so it checks uniformity and independence together.

\subsection{Scoring rules}
\label{sec:eval-scores}
We compare the forecasts with four proper scoring rules, each averaged over the evaluation sample: the log score and the CRPS defined in Section~\ref{sec:intro}, for the whole distribution, and the interval score and the censored likelihood score for central intervals and the tails. The log score and the censored likelihood are positively oriented, and the CRPS and the interval score negatively oriented. By \eqref{eq:expgain}, the expected log-score gain is bounded by the divergence of the PIT distribution from uniformity, so a small gain may reflect how little miscalibration there was to remove rather than a weakness of the method.\footnote{For each score we compare the recalibrated forecast with the raw pool through the per-period difference $d_t = \pm\{S(\tilde F_t, y_t) - S(F_t, y_t)\}$, with the sign chosen so that $d_t > 0$ means the recalibrated forecast did better. In the applications we test $\mathrm{E}[d_t] = 0$ with the unconditional test of \citet{Giacomini2006}, using a Newey--West estimate of the long-run variance \citep{Newey1987} with $\lfloor 4(T/100)^{2/9} \rfloor$ lags. In the simulations, where the ten replications are independent, we use a $t$-test on the per-replication mean gains. Stars denote significance at 10, 5 and 1\%.}

\paragraph{Interval score}
Forecast users often act on central prediction intervals rather than on the full predictive density. For the central $(1-p)\times100\%$ interval with endpoints $q^-_t = F_t^{-1}(p/2)$ and $q^+_t = F_t^{-1}(1-p/2)$, the interval score of \citet{Gneiting2007} is
\begin{equation*}
\mathrm{IS}_p(q^-_t, q^+_t; y_t) = (q^+_t - q^-_t) + \frac{2}{p}(q^-_t - y_t)\,\mathbbm{1}\{y_t < q^-_t\} + \frac{2}{p}(y_t - q^+_t)\,\mathbbm{1}\{y_t > q^+_t\},
\label{eq:is}
\end{equation*}
which rewards narrow intervals and penalises outcomes outside them. We report the 90\% and 95\% intervals, $\mathrm{IS}_{90}$ and $\mathrm{IS}_{95}$.

\paragraph{Censored likelihood score}
To evaluate the tails separately, we use the censored likelihood (CSL) score of \citet{Diks2011}, which \citet{dePunder2026} show is the natural way to localise a strictly proper score while preserving strict propriety. For a region $A_t$,
\begin{equation}
\mathrm{CSL}(F_t, y_t; A_t) = \mathbbm{1}\{y_t \in A_t\} \log f_t(y_t) + \mathbbm{1}\{y_t \notin A_t\} \log\big(1 - F_t(A_t)\big),
\label{eq:csl}
\end{equation}
where $F_t(A_t)$ is the predictive probability of $A_t$. We apply \eqref{eq:csl} to the left tail $A_t = (-\infty, c^{L}_t)$ and the right tail $A_t = (c^{R}_t, \infty)$, giving $\mathrm{CSL}_L$ and $\mathrm{CSL}_R$. Comparing the two shows whether a recalibration improves both tails or only moves probability from one to the other. The thresholds $c^{L}_t$ and $c^{R}_t$ are lower and upper empirical quantiles of a rolling window of past outcomes, so they follow the level of the series, and the tail level is chosen so that each tail contains enough outcomes to evaluate.

\subsection{Method comparison and implementation}
\label{subsec:methods}

\paragraph{The pool}
In our main analysis, the weights $w_{n,t}$ in \eqref{eq:linear_pool} are produced by the LDF of \citet{Bernaciak2024}. The LDF builds an adaptive combination through two layers, each of which discounts past performance. Let $\mathcal{A} = \{\alpha_1,\ldots,\alpha_M\}$ be a grid of discount factors with $\alpha_m \in (0,1]$. In the first layer, for each $\alpha_m$, the models are scored by their log discounted predictive likelihood (LDPL) and weighted by a softmax,
\begin{equation*}
  \mathrm{LDPL}_{n,t}(m) = \sum_{i=1}^{t-1} \alpha_m^{\,i-1} \log f_{n,t-i}(y_{t-i}),
  \qquad
  \omega^{(m)}_{n,t} = \frac{\exp\bigl(\mathrm{LDPL}_{n,t}(m)\bigr)}{\sum_{j=1}^{N}\exp\bigl(\mathrm{LDPL}_{j,t}(m)\bigr)},
  \label{eq:ldf}
\end{equation*}
which gives $M$ pools $F^{(m)}_t = \sum_{n=1}^N \omega^{(m)}_{n,t} F_{n,t}$. A value of $\alpha_m$ close to one rewards performance over a long history, and $\alpha_m = 1$ recovers BMA with equal prior model probabilities, whereas $\alpha_m \rightarrow 0$ favours the most recent performance. The second layer combines these $M$ pools in the same way, with a second grid $\mathcal{A}_2 = \{\alpha^{(2)}_1,\ldots,\alpha^{(2)}_J\}$. For each $\alpha^{(2)}_j$, the pools are scored by their LDPL with discount factor $\alpha^{(2)}_j$ and weighted by a softmax, giving weights $\lambda^{(j)}_{m,t}$ and the forecast
\begin{equation*}
  F^{(j)}_t = \sum_{m=1}^{M} \lambda^{(j)}_{m,t}\, F^{(m)}_t .
\end{equation*}
The second layer therefore chooses how much of the past to use, as the first layer chooses between models. We denote this two-layer LDF with softmax weights by $\mathrm{LDF}^2_{s,s}$, and use $\mathcal{A} = \mathcal{A}_2 = \{1,\allowbreak 0.99,\allowbreak 0.95,\allowbreak 0.9,\allowbreak 0.8,\allowbreak 0.7,\allowbreak 0.6,\allowbreak 0.5,\allowbreak 0.4,\allowbreak 0.3,\allowbreak 0.2,\allowbreak 0.001\}$.\footnote{Results labelled $\alpha = 0.9$ use the second-layer discount factor $\alpha^{(2)}_j = 0.9$.}

Two properties of this construction matter for recalibration. First, $F^{(j)}_t$ is again a linear pool, with weights $w_{n,t} = \sum_{m} \lambda^{(j)}_{m,t}\, \omega^{(m)}_{n,t}$, so it is subject to the miscalibration described in Section~\ref{sec:intro}. Second, every weight depends only on outcomes up to $t-1$, so $F_t$ is a genuine one-step-ahead forecast. The recalibration methods take $F_t$ as given and leave its weights unchanged, so they apply equally to any other weighting scheme.

\paragraph{Static maps}
We compare six recalibration methods, the first two of which are static. If the miscalibration does not change over time, $\Pi_t = \Pi$, then $\theta^{*}_t = \theta^{*}$ is constant, and a natural estimator maximises the average gain over all past PITs,
\begin{equation*}
  \hat\theta^{\mathrm{S}}_{t} = \arg\max_{\theta}\; \frac{1}{t_r-1}\sum_{s=1}^{t_r-1} \log \tau'_{\theta}(u_s) \;-\; \lambda \lVert \theta \rVert^2,
  \qquad t_r = R\lfloor (t-1)/R \rfloor + 1,
  \label{eq:static}
\end{equation*}
where $t_r$ is the most recent refit date. The map is therefore re-estimated on an expanding window every $R$ periods and held fixed in between. The ridge term, with $\lambda = 0.01$, shrinks the estimate towards the identity map $\theta = 0$ and stabilises the fit when few PITs are available. With the Beta family and $\lambda = 0$, this is the BLP of \citet{Gneiting2013}, estimated recursively. The static map serves both as a recalibration method and as the benchmark for the time-varying methods.

\paragraph{Time-varying maps}
The other four methods cross the two families with the two updating schemes of Section~\ref{sec:method}: Beta-GAS, RQS-GAS, Beta-Con and RQS-Con. All four use the filter rates in Table~\ref{tab:experts} and the same DMA settings, $\alpha^{\mathrm{DMA}} = 0.95$ and $c_\omega = 10^{-3}$, and every filter starts at the identity map. The network of the convolution is trained by Adam with learning rate $10^{-3}$ for at most 300 epochs.

\section{Simulation Study}
\label{sec:sim}
\paragraph{Benchmark design}
We start from the DGP of \citet{Diebold2023}, as extended by \citet{Bernaciak2024}, which was designed to evaluate adaptive combination methods under unobserved level shifts. The observed variable $y_t$ follows
\begin{align*}
  y_t &= \mu_t + \phi_x x_t + \sigma_y \varepsilon_t, & \varepsilon_t &\overset{\text{iid}}{\sim} \mathcal{N}(0,1), \label{eq:dgp-y}\\
  x_t &= \phi_x x_{t-1} + \sigma_x e_t, & e_t &\overset{\text{iid}}{\sim} \mathcal{N}(0,1), 
\end{align*}
where $x_t$ is a long-run component of $y_t$ and $\mu_t \in \{-1, 0, 1\} $ is a time-varying level. Each forecaster observes the DGP structure but not the regime state $\mu_t$, so the $n$th forecaster predicts from a noisy observation of $x_t$:
\begin{align*}
z_{nt} &= x_t + \sigma_{z}\,\nu_{nt}, 
& \nu_{nt} &\sim \mathcal{N}(0,1),\\
\tilde y_{nt} &= \eta_n + z_{nt} + \sigma_y \,\varepsilon_t, 
& \varepsilon_t &\sim \mathcal{N}(0,1).
\end{align*}
We set $\phi_x = 0.9$, $\sigma_x = 0.5$, $\sigma_y = 0.5$, $\sigma_{z} = 0.7$, 
$N = 20$, $T = 2001$, $\eta_n = -2 + 0.2105\,(n-1), \quad n=1,\ldots,N$.

\begin{figure}[H]
    \centering
    \includegraphics[width=\linewidth]{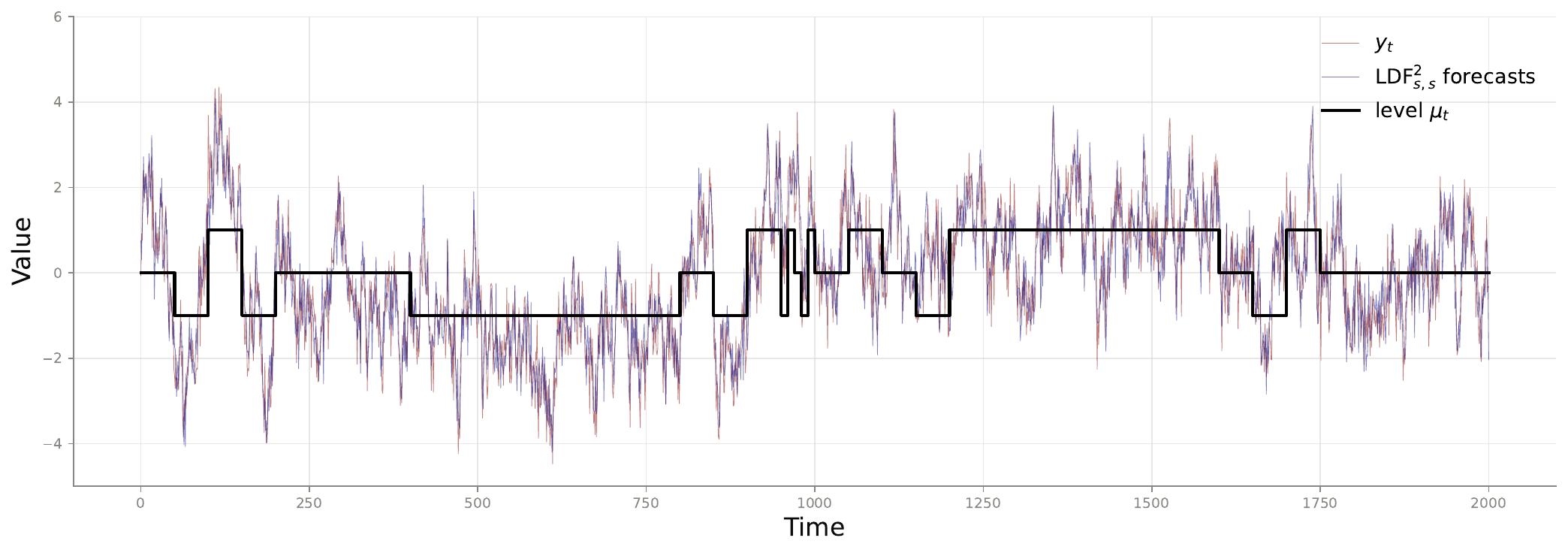}
    \caption{Simulated series $y_t$ and $\mathrm{LDF}^2_{s,s}$ forecasts at $\alpha=0.9$ in the over-dispersed scenario.}
    \label{fig:simy}
\end{figure}

We consider two specifications of $\mu_t$. In the first, following \citet{Bernaciak2024}, $\mu_t$ is the deterministic piecewise-constant path
\begin{equation}
\label{eq:mu_t}
\footnotesize
\mu_t =
\begin{cases}
0, & \text{for } t \in
\begin{aligned}[t]
&[0,49]\ \cup\ [200,399]\ \cup\ [800,849]\ \cup\ [970,979]
\cup\ [1000,1049]\ \\ \ &\cup\ [1600,1650]\ \cup\ [1750,2001],
\end{aligned}\\[6pt]
1, & \text{for } t \in
\begin{aligned}[t]
&[100,150]\ \cup\ [900,949]\ \cup\ [960,969]\ \cup\ [990,999] \cup\ [1050,1099]\ \\ \ &\cup\ [1200,1599]\ \cup\ [1700,1749],
\end{aligned}\\[8pt]
-1, & \text{otherwise.}
\end{cases}
\end{equation}
This specification produces an over-dispersed pool, which we take as our first scenario and use as the basis for constructing the remaining scenarios. Figure \ref{fig:simy} displays the forecasts alongside the simulated series $y_t$. In the second specification, $\mu_t$ follows a time-varying Markov chain on $\{-1, 0, 1\}$, started at $\mu_0 = 0$, with transition matrix $P^{(1)}$ for $t < 1000$ and $P^{(2)}$ for $t \geq 1000$,
\begin{equation*}
P^{(1)} =
\begin{bmatrix}
0.99 & 0.005 & 0.005 \\
0.005 & 0.99 & 0.005 \\
0.005 & 0.005 & 0.99
\end{bmatrix},
\qquad
P^{(2)} =
\begin{bmatrix}
0.98 & 0.01 & 0.01 \\
0.01 & 0.98 & 0.01 \\
0.01 & 0.01 & 0.98
\end{bmatrix}.
\label{eq:markov}
\end{equation*}

\begin{figure}[H]
    \centering
    \includegraphics[width=\linewidth]{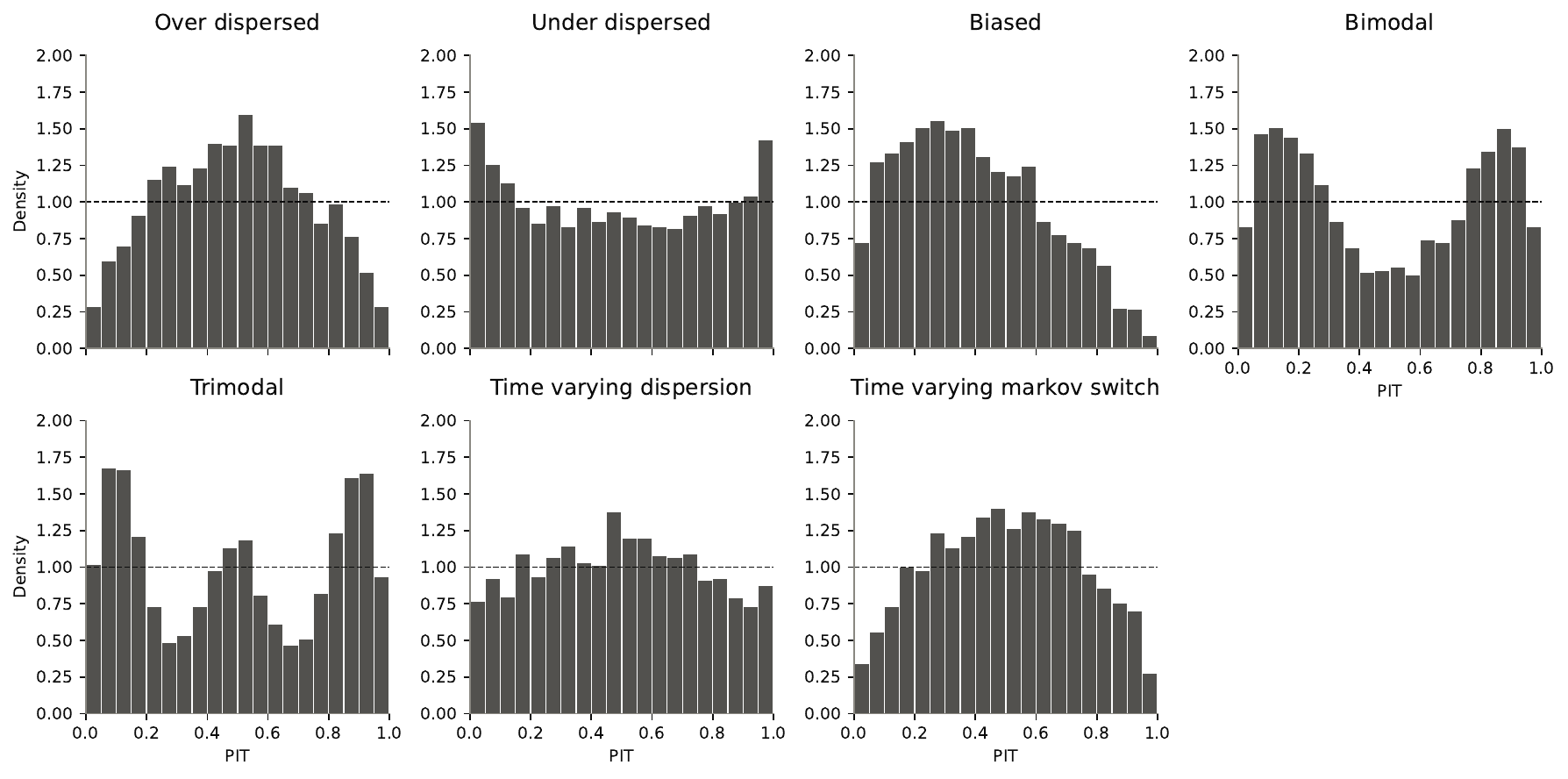}
    \caption{\footnotesize Raw pool PIT histograms for all simulation scenarios with $\alpha=0.9$.}
    \label{fig:raw-pit}
\end{figure}

\begin{figure}[H]
    \centering
    \includegraphics[width=\linewidth]{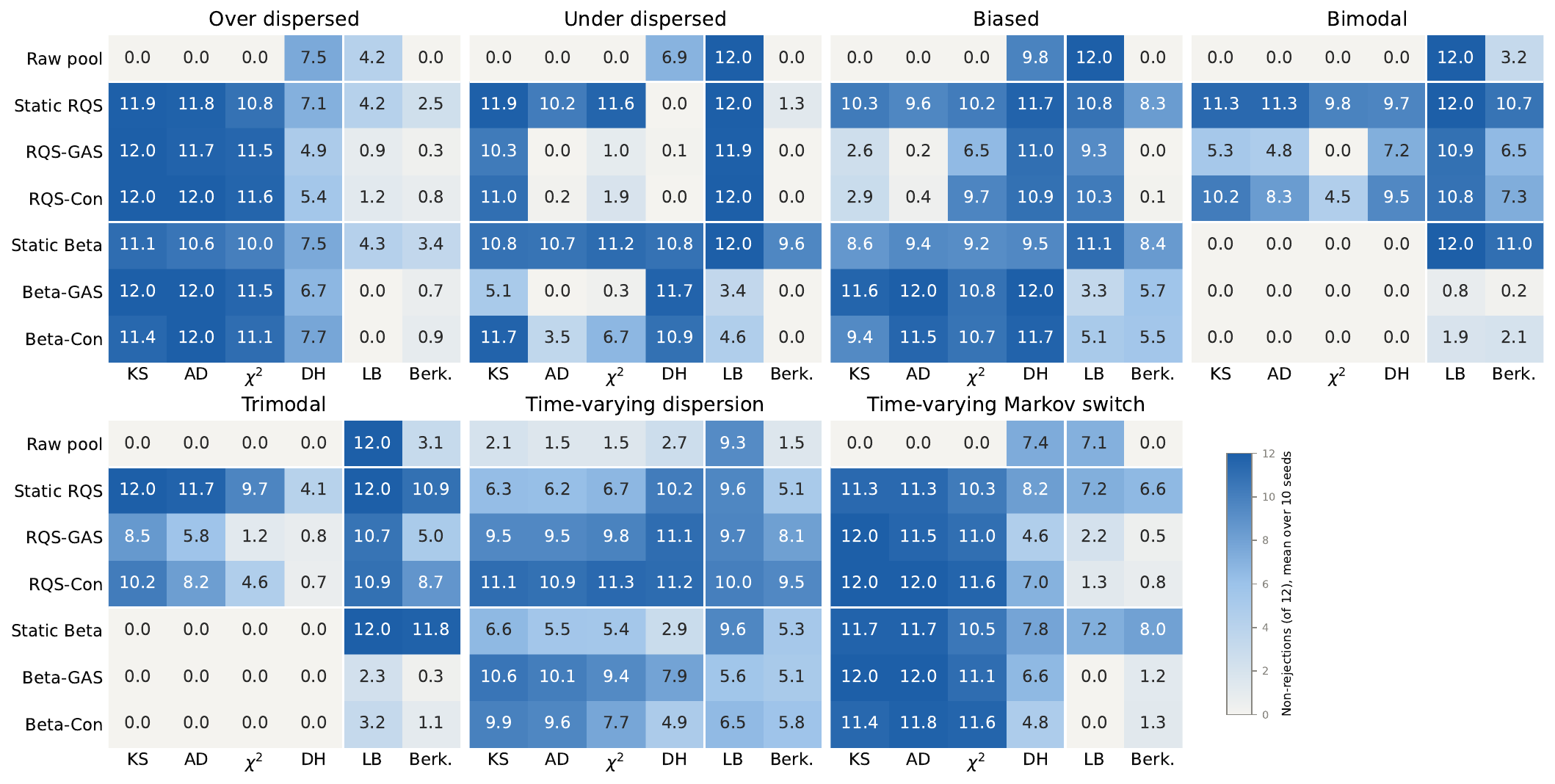}
    \caption{\footnotesize Mean number of the 12 LDF discount factors at which each test does not reject at 5\%, over ten replications; darker is better.}
    \label{fig:cal-tests}
\end{figure}

\paragraph{Modified designs}
The level $\mu_t$ follows \eqref{eq:mu_t}, but the
forecasters now observe the conditional mean
$m_t = \mu_t + \phi_x x_t$ and produce point forecasts
\begin{equation}
  \hat y_{nt} = m_t + \eta_n + \varsigma_t \nu_{nt},
  \qquad \nu_{nt} \overset{\mathrm{iid}}{\sim}
  \mathcal{N}(0,1).
  \label{eq:sim-baseline}
\end{equation}
Each forecaster issues a Gaussian predictive distribution
$F_{n,t} = \mathcal{N}(\hat y_{nt},\sigma_y^2)$,
so $\varsigma_t$ controls the dispersion of the
predictive means across forecasters, while the
variance of each component forecast remains fixed
at $\sigma_y^2$. With $\eta_n=0$ and
$\varsigma_t=\sigma_y$, the predictive means
are centred on the true conditional mean.
Each scenario modifies this baseline to introduce
a particular form of miscalibration:
\begin{itemize}
\item \textbf{Under-dispersed.} The target noise is wider than the forecasters assume, $y_t = m_t + 1.5\,\sigma_y \varepsilon_t$, with $\varsigma_t = 0.9\sigma_y$. 
\item \textbf{Biased.} All forecasters share a common bias, $\eta_n = 0.2$. 
\item \textbf{Bimodal.} The target receives a symmetric jump, $y_t = m_t + J_t + \sigma_y \varepsilon_t$ with $J_t \overset{\text{iid}}{\sim} \mathrm{Uniform}\{-1, 1\}$, with $\varsigma_t = (\sigma_y^2 + 1)^{1/2}$. 
\item \textbf{Trimodal.} The target is concentrated around three points, $y_t = m_t + J_t + 0.15\,\varepsilon_t$ with $J_t \overset{\text{iid}}{\sim} \mathrm{Uniform}\{-1.5, 0, 1.5\}$, and $\varsigma_t = (0.15^2 + 1.5)^{1/2}$. 
\item \textbf{Time-varying dispersion.} A Markov chain $S_t \in \{-1, 0, 1\}$ with transition matrix $P^{(1)}$ switches the forecast spread between $\varsigma_t = 0.05\,\sigma_y$ when $S_t \neq 0$ and $\varsigma_t = 5\,\sigma_y$ when $S_t = 0$.
\end{itemize}
We recalibrate each scenario with the six methods of Section~\ref{subsec:methods}. The RQS uses $K = 5$ bins, the neural network has $H = 4$ hidden units and is refitted every $R = 50$ periods. Each scenario is replicated ten times, with the level path fixed and the random draws regenerated. Figure~\ref{fig:raw-pit} shows the PIT histograms of the raw $\mathrm{LDF}^2_{s,s}$ pool in each scenario, which reveal the form of miscalibration. In the over-dispersed and time-varying Markov switch scenarios the histogram is hump-shaped, because the pool is too wide. The under-dispersed pool is too narrow and gives a U-shaped histogram, with excess mass near 0 and 1. The common bias tilts the histogram, moving mass towards one end of the unit interval. The bimodal and trimodal scenarios produce two and three peaks. Under time-varying dispersion the histogram is close to flat, because the narrow and wide regimes produce opposite distortions that largely cancel when pooled over the sample.
The multimodal scenarios show that recalibration can restore uniformity only if the map family matches the shape of the PIT distribution. In the bimodal and trimodal scenarios, no Beta method passes KS, AD, $\chi^2$ or DH at any discount factor, whether the map is static or adaptive (Figure~\ref{fig:cal-tests}), and the recalibrated ECDFs remain close to that of the raw pool (Figure~\ref{fig:sim-ecdf}). The RQS maps, by contrast, remove most of the distortion. The Static-RQS map passes most uniformity tests in both scenarios, and RQS-Con passes KS and AD at most discount factors. The adaptive RQS maps still often fail $\chi^2$, and DH in the trimodal scenario, which suggests they leave small local departures near the original modes.

In the over-dispersed and time-varying Markov switch scenarios, the two map families perform similarly, and every map passes KS, AD and $\chi^2$ at ten or more discount factors. The Beta family has the advantage when the distortion lies in the tails. In the under-dispersed scenario, every RQS map fails DH, whereas all three Beta maps pass it. The Static-Beta map passes every uniformity test at almost every discount factor. Under the common bias, the Beta maps pass all four uniformity tests at most discount factors, while the adaptive RQS maps fail KS and AD. A plausible reason is that a Beta density can become unbounded at 0 and 1, whereas the derivative of the spline at the end knots is finite, which limits how much tail mass it can capture. Under time-varying dispersion, the adaptive maps clearly outperform the static ones: RQS-Con passes the four uniformity tests at about 11 of 12 discount factors, against about 6 for both static maps.

This advantage of the adaptive maps comes at a cost in independence. Because the parameter of an adaptive map is driven by past PITs, the recalibrated PITs can inherit serial dependence from the filter. The adaptive Beta maps lower the number of discount factors at which the LB test does not reject from 7.1 to 0.0 in the time-varying Markov switch scenario. The adaptive RQS maps lower it less, and the static maps leave it essentially unchanged. Adaptive updating can therefore restore uniformity at the expense of adding serial dependence to the PITs.
The time-varying methods can improve on a static map only when the miscalibration changes over time. In the under-dispersed and biased scenarios the miscalibration is constant. The forecasters observe $m_t$, so a level switch shifts every component equally and leaves the shape of the pool unchanged. In both scenarios the static map is the best method in each family: GAS falls behind it by 0.015--0.021 per period, and the convolution by 0.015--0.022 (Table~\ref{tab:sim-pairwise}). When the optimal map is constant, the static map, re-estimated on an expanding window, uses all past PITs and becomes more precise over time. An adaptive map, whose updates have a constant step size, gives most weight to recent PITs and keeps fluctuating around the optimal map.

GAS falls behind the static map in six of the seven scenarios (Table~\ref{tab:sim-pairwise}). This includes the over-dispersed and time-varying Markov switch scenarios, in which the miscalibration does change. In these scenarios, however, much of the miscalibration is a persistent over-dispersion that the static map already captures, and the remainder drifts gradually as the weights adjust after each switch. A filter driven by past PITs learns about this drift only once it appears in the PITs, so it keeps chasing a moving target. The time-varying dispersion scenario is the exception. There, GAS outperforms the static map in every replication, by 0.072 per period for RQS and 0.091 for Beta. The two regimes produce opposite distortions that largely cancel over the sample, so the static map learns a correction close to the identity. The miscalibration within each regime, by contrast, is large and stable enough for the filter to learn.

The convolution generally improves on GAS in five of the seven scenarios, although the magnitude of the improvement depends on the map family. The convolution falls behind the static map only in the under-dispersed and biased scenarios, where the shape of the pool does not change.

\begin{table}[t]\centering\footnotesize
\begin{tabular}{llccc}\toprule
Scenario & Family & Con $-$ GAS & GAS $-$ static & Con $-$ static \\ \midrule
Over-dispersed & RQS & $+0.056^{***}$ (10/10) & $-0.018^{***}$ (0/10) & $+0.038^{***}$ (10/10) \\
 & Beta & $+0.075^{***}$ (10/10) & $-0.016^{***}$ (0/10) & $+0.059^{***}$ (10/10) \\
\addlinespace
Under-dispersed & RQS & $-0.004^{***}$ (1/10) & $-0.018^{***}$ (0/10) & $-0.022^{***}$ (0/10) \\
 & Beta & $+0.006^{***}$ (10/10) & $-0.021^{***}$ (0/10) & $-0.016^{***}$ (0/10) \\
\addlinespace
Biased & RQS & $-0.003^{*}$ (2/10) & $-0.018^{***}$ (0/10) & $-0.022^{***}$ (0/10) \\
 & Beta & $+0.000$ (5/10) & $-0.015^{***}$ (0/10) & $-0.015^{***}$ (0/10) \\
\addlinespace
Bimodal & RQS & $+0.020^{***}$ (10/10) & $-0.016^{***}$ (0/10) & $+0.004$ (7/10) \\
 & Beta & $+0.013^{***}$ (10/10) & $-0.011^{***}$ (0/10) & $+0.002$ (5/10) \\
\addlinespace
Trimodal & RQS & $+0.058^{***}$ (10/10) & $-0.022^{***}$ (0/10) & $+0.036^{***}$ (10/10) \\
 & Beta & $+0.017^{***}$ (10/10) & $-0.012^{***}$ (0/10) & $+0.005^{***}$ (8/10) \\
\addlinespace
Time-varying & RQS & $+0.022^{**}$ (10/10) & $+0.072^{***}$ (10/10) & $+0.094^{***}$ (10/10) \\
 dispersion & Beta & $+0.027^{***}$ (10/10) & $+0.091^{***}$ (10/10) & $+0.118^{***}$ (10/10) \\
\addlinespace
Time-varying  & RQS & $+0.052^{***}$ (10/10) & $-0.017^{***}$ (0/10) & $+0.035^{***}$ (10/10) \\
Markov switch & Beta & $+0.066^{***}$ (10/10) & $-0.014^{***}$ (0/10) & $+0.052^{***}$ (10/10) \\
\bottomrule\end{tabular}
\caption{Mean log-score difference per period at $\alpha=0.9$ for 10 replications. Stars: $t$-test across replications at 10/5/1\%; in brackets, replications in which the first method is ahead.}
\label{tab:sim-pairwise}
\end{table}

\begin{figure}[t]
  \centering
  \includegraphics[width=\textwidth]{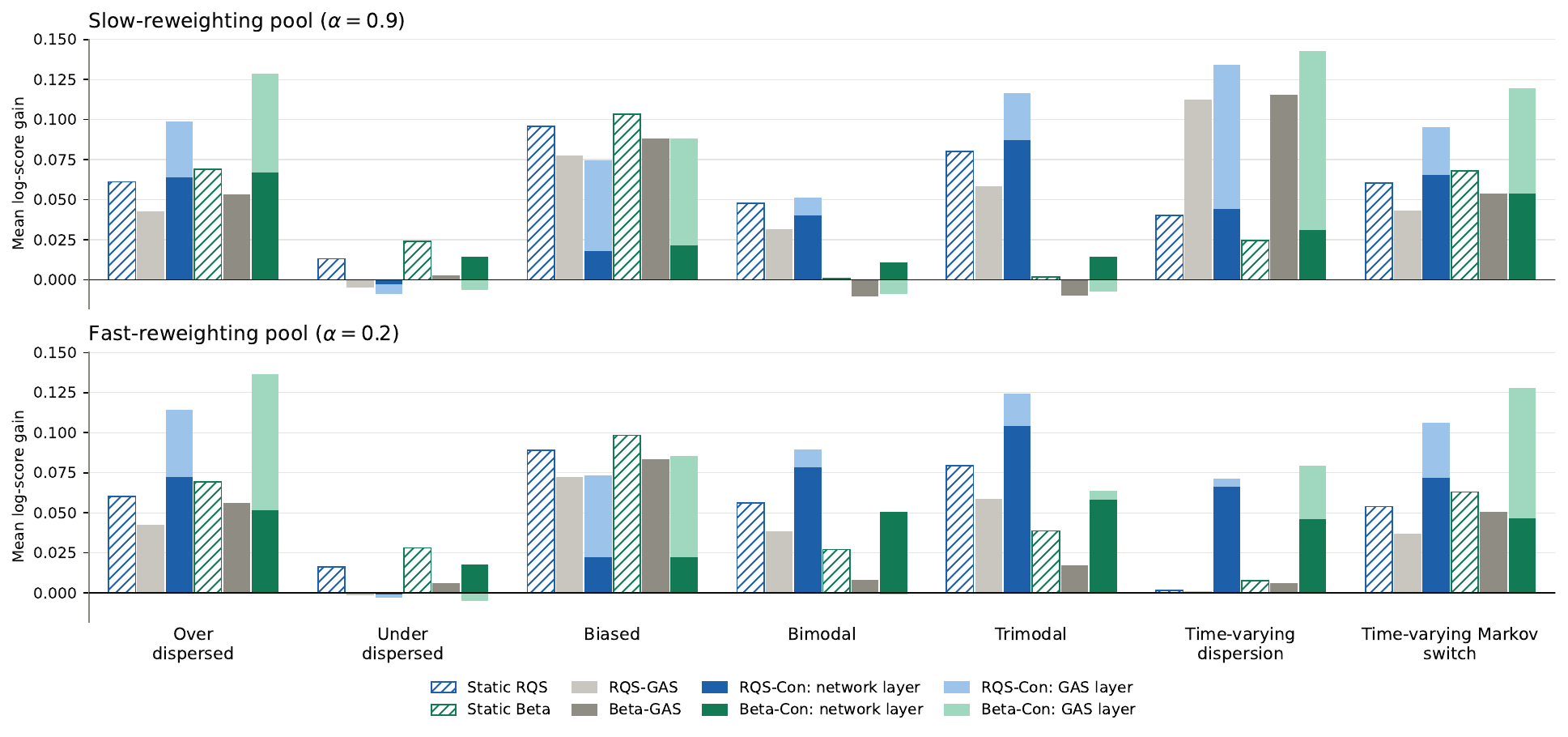}
  \caption{\footnotesize Mean log-score gain over the raw pool per period, over ten replications.  Grey: GAS. Coloured: the convolution, split into its network layer (dark) and its GAS layer (light); blue for RQS, green for Beta. Hatched: static map of the same family. Top: $\alpha = 0.9$; bottom: $\alpha = 0.2$.}
  \label{fig:sim_layers}
\end{figure}

The convolution's log-score gain splits into a network component and a GAS component. Figure~\ref{fig:sim_layers} reports this split, for a pool that reweights slowly ($\alpha = 0.9$) and one that reweights quickly ($\alpha = 0.2$). When the pool reweights slowly, most of the gain comes from the GAS layer. Under time-varying dispersion at $\alpha = 0.9$, GAS alone gains about 0.11 per period in both families, far more than the static map. When the pool reweights quickly, its miscalibration changes from one period to the next, and neither the static map nor the GAS filter can follow it. At $\alpha = 0.2$, their gains under time-varying dispersion fall to around zero in both families. The network layer, by contrast, still gains about 0.065 (RQS) and 0.045 (Beta) per period.

\begin{figure}[H]
    \centering
    \includegraphics[width=\linewidth]{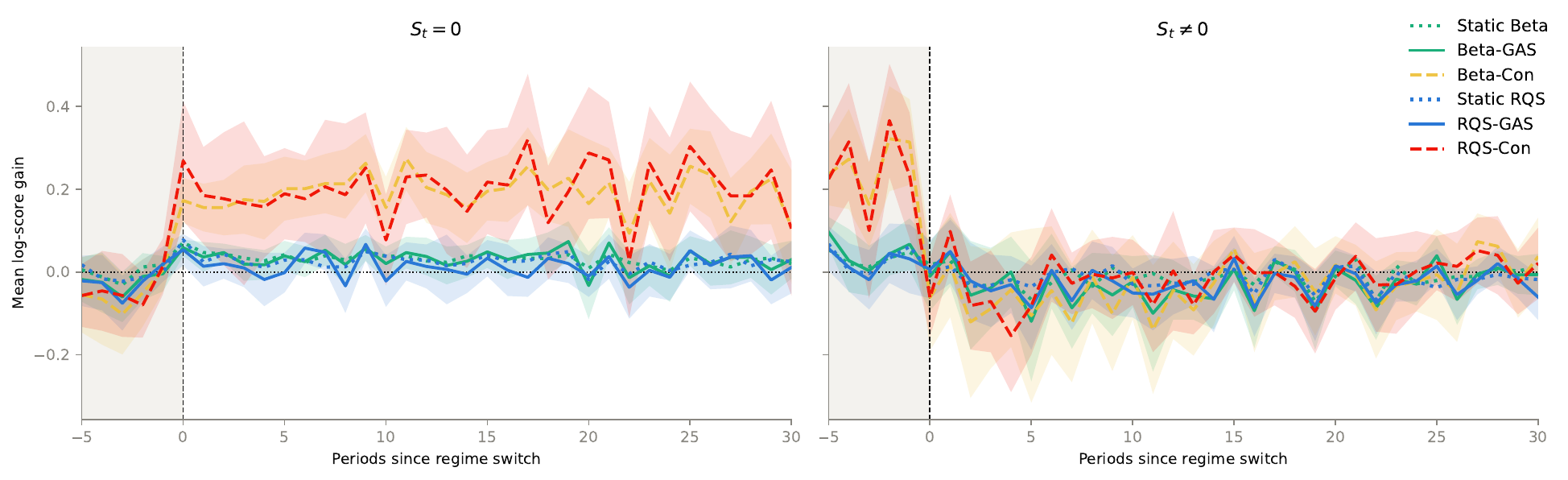}
    \caption{Per-period mean log-score gain over the raw pool, averaged over the regime switches (pooled across replications) and aligned so that lag 0 is the first period of the new regime. Left: switches into the dispersed regime $S_t = 0$; right: switches back to $S_t \neq 0$. Shaded bands show empirical 95\% intervals across regime switches.}
    \label{fig:event-study}
\end{figure}

Figure~\ref{fig:event-study} shows the same mechanism over time. It aligns all regime switches of the time-varying dispersion scenario, and averages the per-period log-score gain around each switch. When the members start to disperse ($S_t = 0$), the gain of both convolution methods jumps to about 0.2 per period in the first period of the new regime and stays there. GAS and the static maps remain close to zero for at least 30 periods. Since a filter needs several periods to learn a new correction, the immediate jump comes from the network, which detects the widening of the pool as soon as it occurs. When the members agree again ($S_t \neq 0$), the gain of the convolution falls and stays close to zero like the other methods.

\section{Empirical applications}
\label{sec:empirical}
We apply the recalibration methods to nine G10 exchange rates and US inflation. In each application we also run the dynamic BPS of \citet{McAlinn2019} on the same agents and apply the recalibration methods to its predictive distributions. The LDF combines the agents' densities without modelling their biases or dispersion, leaving these to be corrected afterwards; BPS learns them inside the synthesis, through a time-varying intercept and agent coefficients. The comparison asks whether recalibration still helps a combination that models the agents' miscalibration itself. BPS needs a training window of 50 periods, so in each application it is compared with the LDF on a shorter evaluation window (detailed configurations in~\ref{app:bps}).

\subsection{Exchange rates}
\label{sec:ex}
Following \citet{Bernaciak2024}, we forecast the monthly returns of nine currencies against the US dollar (USD): Australian dollar (AUD), Canadian dollar (CAD), euro (EUR), Japanese yen (JPY), New Zealand dollar (NZD), Norwegian krone (NOK), Swedish krona (SEK), Swiss franc (CHF) and pound sterling (GBP). The data run from June 1990 to July 2020. The pool combines 32 time-varying parameter Bayesian vector auto-regressive (TVP-BVAR) models of \citet{Beckmann2020}, each using the uncovered interest parity (UIP) fundamental as the predictor. The coefficient forgetting factor is set to one, so the coefficients are constant over time, and the construction is described in~\ref{app:tvpvar}. The RQS map uses $K=4$ bins, and the neural network has $H=4$ hidden units, refitted every $R=80$ months.

The raw LDF pool is close to calibrated for most currencies. The KS test does not reject it for any of the nine (Table~\ref{tab:fx-scores}), and the pool passes at least five of the six calibration tests for six of them (Table~\ref{tab:fx-tests}). The exceptions are JPY, which passes only two tests, and CHF and GBP, which pass four. Every recalibration method raises JPY to five passed tests, whereas for the other currencies recalibration changes the test outcomes little, as there is little miscalibration to correct.
\begin{table}[H]\centering\small
\begin{tabular}{lccccccccc}\toprule
Method & AUD & CAD & EUR & JPY & NZD & NOK & SEK & CHF & GBP \\ \midrule
Raw pool & 5 & 5 & 5 & 2 & 6 & 5 & 6 & 4 & 4 \\
\midrule
Static-RQS & 6 & 5 & 5 & 5 & 6 & 6 & 6 & 5 & 6 \\
RQS-GAS & 6 & 5 & 5 & 5 & 6 & 6 & 6 & 5 & 4 \\
RQS-Con & 6 & 5 & 5 & 5 & 6 & 6 & 6 & 5 & 5 \\
\midrule
Static-Beta & 6 & 5 & 5 & 5 & 6 & 5 & 6 & 5 & 5 \\
Beta-GAS & 6 & 5 & 5 & 5 & 6 & 6 & 6 & 4 & 4 \\
Beta-Con & 6 & 6 & 5 & 5 & 6 & 6 & 6 & 5 & 4 \\
\bottomrule\end{tabular}
\caption{Exchange rates, $\alpha=0.9$: number of calibration tests (KS, AD, $\chi^2$, DH, LB and Berkowitz) that do not reject at 5\% for each currency.}
\label{tab:fx-tests}
\end{table}

\begin{figure}[H]
    \centering
    \includegraphics[width=\linewidth]{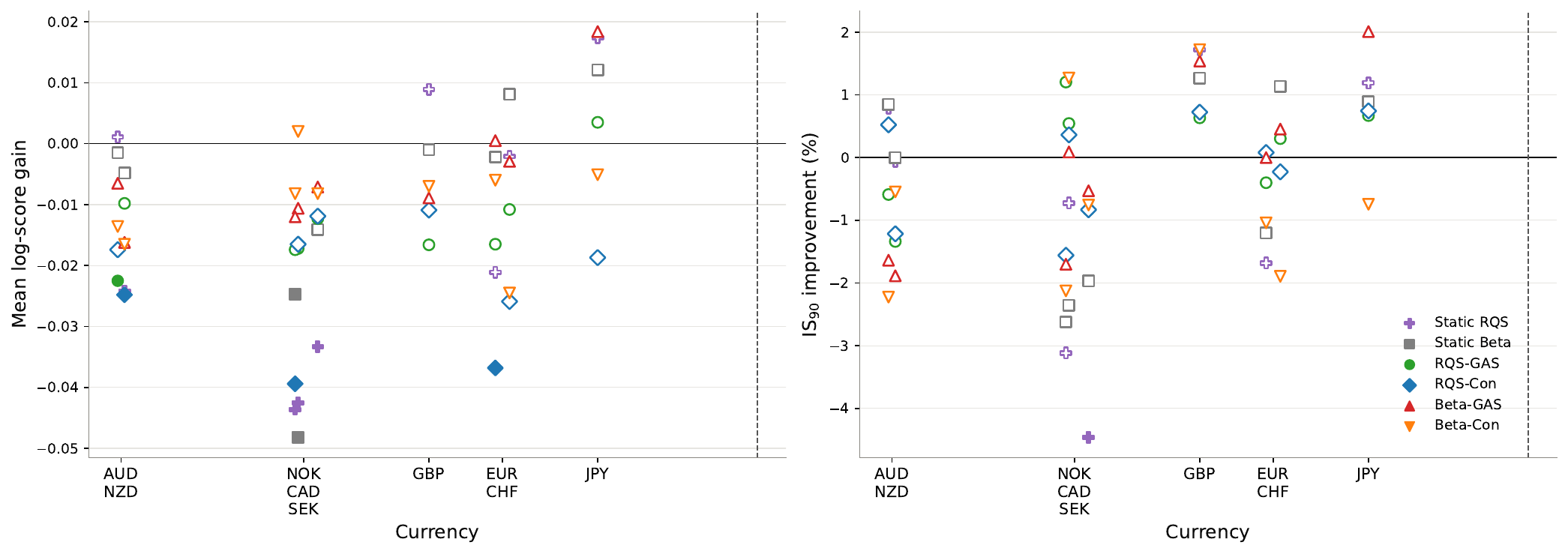}
    \caption{Exchange rates, $\alpha = 0.9$: gain of each method over the raw pool against the raw pool's miscalibration for the nine currencies. The currencies are ordered by their raw-pool KS distances. Left: mean log-score gain; right: improvement in $\mathrm{IS}_{90}$ (\%). Filled markers: difference from the raw pool significant at 5\% (HAC Giacomini–White test); open markers: not significant. Dashed line: 5\% critical value of the KS test; on the left side of the line the raw pool is not rejected.}
    \label{fig:fx-miscal}
\end{figure}

Recalibration does not significantly improve the log score of the LDF pool. No method improves its log score significantly at the 5\% level. The significant changes are all losses, and they come from the static maps and the RQS adaptive maps (Table~\ref{tab:fx-scores}). The Static-RQS map, for example, lowers the log score by about 0.04 per month for NOK and CAD. Beta-GAS and Beta-Con are the only methods that never lose significantly. This pattern is consistent with the bound in \eqref{eq:klbound}. When the PIT distribution is close to uniform there is little to gain, and estimation noise can turn the small potential gain into a small loss, especially for maps with many parameters. Accordingly, the only currency for which most methods gain is JPY, the pool with the largest KS distance (Figure~\ref{fig:fx-miscal}).

\begin{figure}[H]
    \centering
    \includegraphics[width=\linewidth]{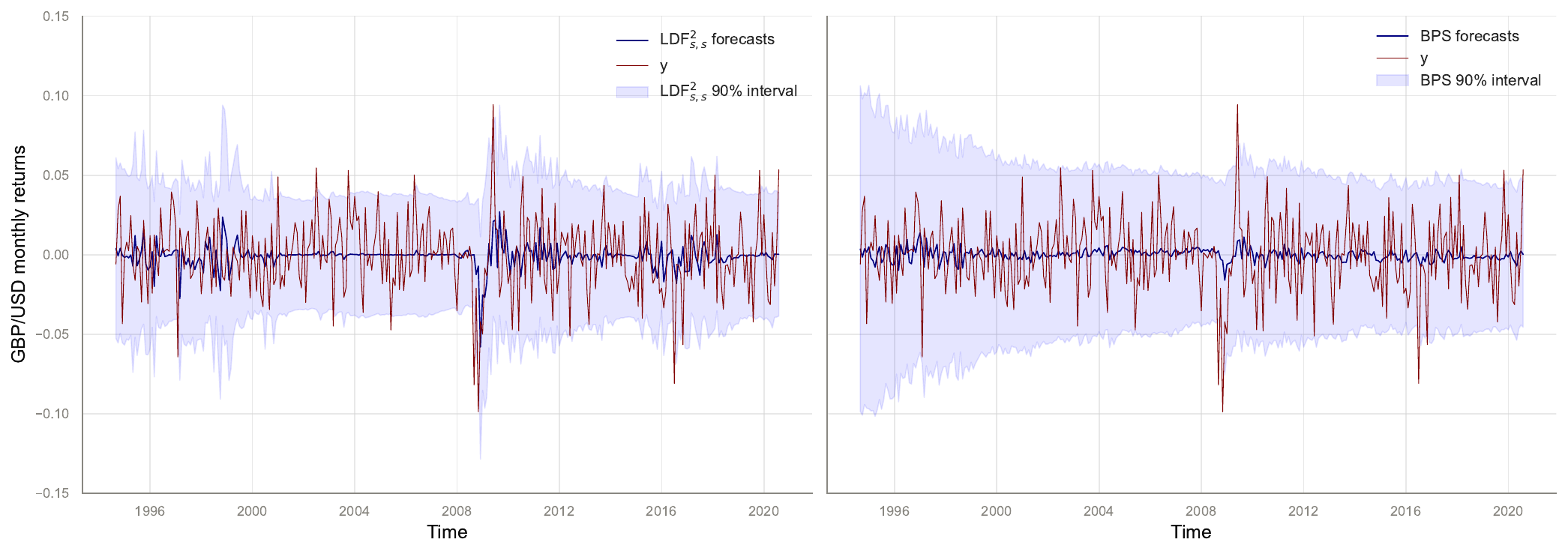}
    \caption{GBP/USD monthly returns (red) with forecasts (blue) and 90\% interval (shaded) of the raw $\mathrm{LDF}^2_{s,s}$ pool (left) and of BPS (right).}
    \label{fig:fx-intervals}
\end{figure}

BPS behaves differently. On the BPS evaluation window, BPS and the LDF pool forecast the mean equally well and have similar mean log scores, but their intervals differ (Table~\ref{tab:raw-pool-comparison}). BPS gives wider 90\% intervals than the LDF pool for every currency, covering up to 96\% of outcomes. Figure~\ref{fig:fx-intervals} illustrates this for GBP: the BPS interval narrows slowly and smoothly and barely reacts to the 2008--09 volatility, whereas the LDF interval widens sharply and then contracts. Given the similar point accuracy and the higher empirical coverage of the BPS intervals, their greater width suggests over-dispersion. The KS distance of BPS exceeds that of the LDF pool for every currency, and the KS test rejects BPS for CHF and EUR (Table~\ref{tab:fx-scores-bps}).

\begin{figure}[H]
    \centering
    \includegraphics[width=\linewidth]{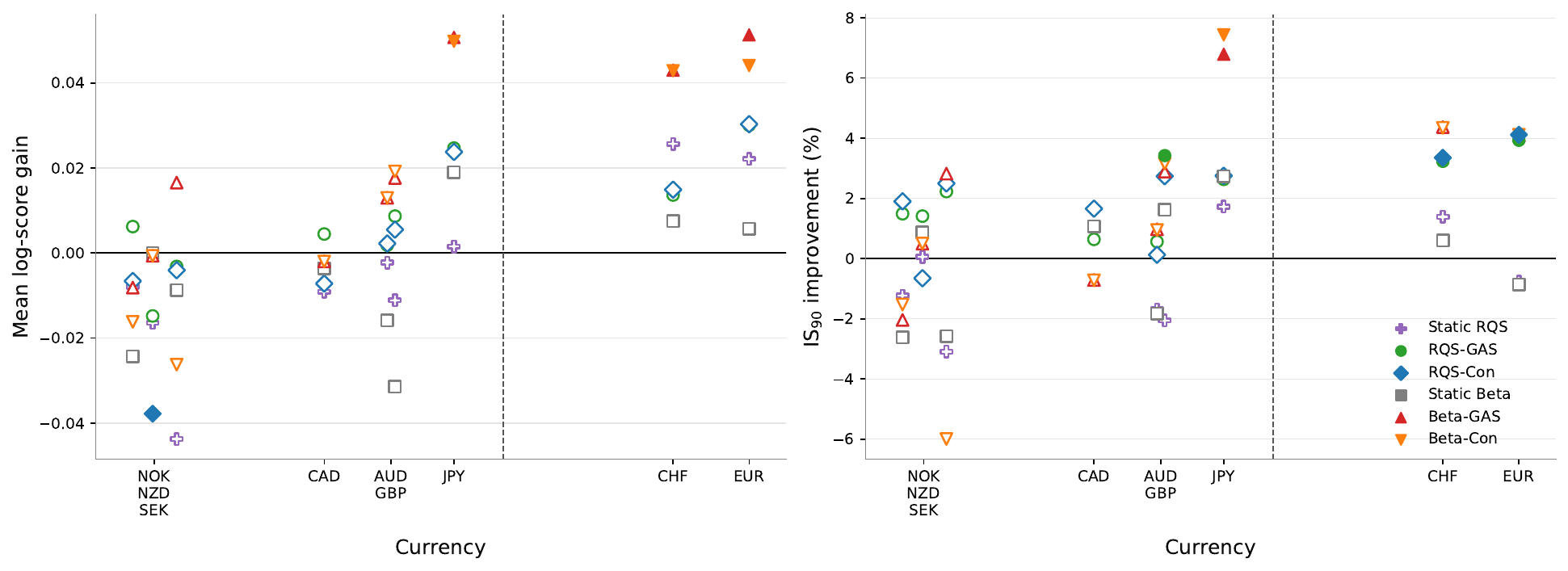}
    \caption{Exchange rate results for BPS. Conventions as in Figure~\ref{fig:fx-miscal}.}
    \label{fig:bps-gain-mis}
\end{figure}

\vspace{-0.13cm}
Because BPS is miscalibrated, it gains from recalibration, and the gains follow its miscalibration (Figure~\ref{fig:bps-gain-mis}). For EUR, CHF and JPY, Beta-GAS and Beta-Con raise the log score significantly, by up to 0.05 per month. The interval score $\mathrm{IS}_{90}$ improves as well, by about 7\% for JPY under the adaptive Beta maps. For the less miscalibrated syntheses, the changes are small and mostly insignificant. The adaptive maps also outperform the static ones for most currencies. One possible explanation, which we do not test formally, is that the over-dispersion of BPS varies over the sample. It is largest early on, when the interval is widest. An adaptive map can therefore shrink the interval more at the start than later, which a single static map cannot.

\subsection{US inflation}
The second application follows \citet{McAlinn2019} in forecasting the quarterly US inflation rate from 1961Q1 to 2014Q4, measured as the annual percentage change in the chain-weighted GDP price index. The pool combines four models: $M_1$ uses the first lag of inflation; $M_2$ the first three lags of inflation, the interest rate and the unemployment rate; $M_3$ the first three lags of inflation only; and $M_4$ the first lag of inflation, the interest rate and the unemployment rate. All four produce Student-$t$ predictive distributions with about 20 degrees of freedom. The RQS map uses $K=4$ bins, and the neural network has $H=6$ hidden units, refitted every $R=10$ quarters. 

\begin{figure}[t]
    \centering
    \includegraphics[width=\linewidth]{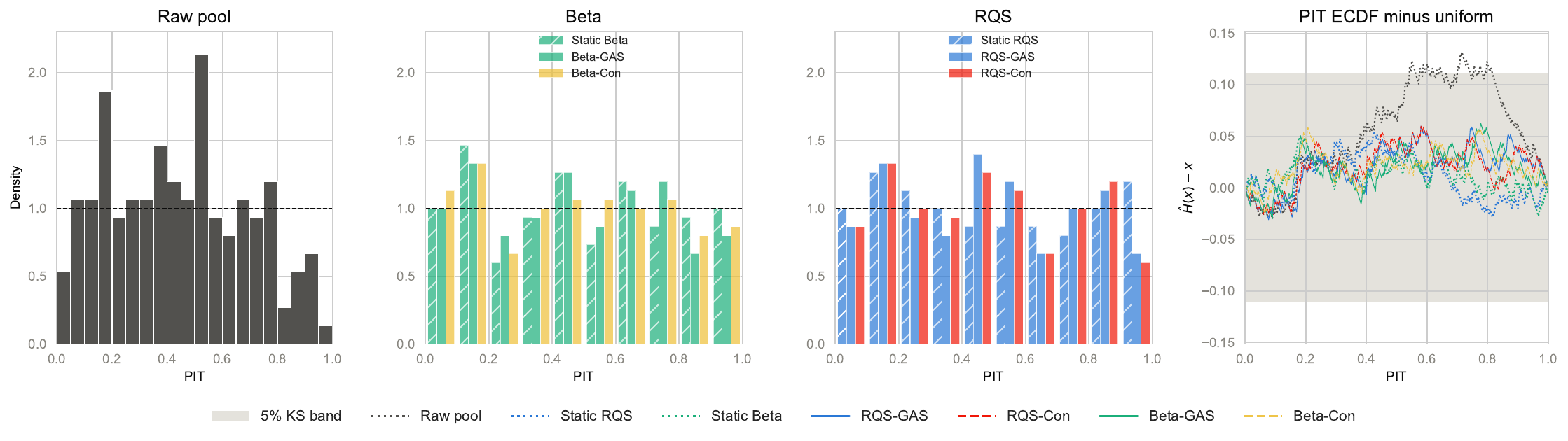}
    \caption{\footnotesize PIT diagnostics for US inflation at $\alpha = 0.9$. From left to right:
PIT histogram of the raw $\mathrm{LDF}^2_{s,s}$ pool; PIT histograms after
recalibration with the Beta and the RQS maps (static, GAS and Con); and the deviation of each empirical PIT CDF from the uniform CDF, $\hat H(x) - x$, with the 5\% KS acceptance band shaded. }
    \label{fig:infl-pit}
\end{figure}

The raw pool fails the KS, AD and $\chi^2$ tests at every discount factor (Table~\ref{tab:inf-tests-passed}), and the deviation of its empirical PIT CDF from the uniform CDF leaves the 5\% KS band (Figure~\ref{fig:infl-pit}). The deviation is positive over most of the unit interval, peaking at PITs around $0.7$, and the PIT histogram has little mass above 0.8. Outcomes therefore fall too often in the lower part of the predictive distribution: the pool tends to over-predict inflation, and its upper tail is too heavy. 

All six recalibration methods pass the four uniformity tests and the Berkowitz test at every discount factor (Table~\ref{tab:inf-tests-passed}). The LB test, however, rejects for every forecast, including the raw pool: the remaining serial dependence may reflect dynamics not captured by the component models or the recalibration maps. The difference between the tests may reflect their sensitivity to different forms of serial dependence, since Berkowitz assesses first-order dependence of the probit-transformed PITs, whereas Ljung–Box jointly assesses autocorrelations of the PITs at lags one to four.

\begin{table}[H]
\centering
\footnotesize
\begin{tabular}{l c c c c c c }
\toprule
 & KS & AD & $\chi^2$ & DH & LB & Berkowitz \\
\midrule
Raw pool & 0 & 0 & 0 & 12 & 0 & 0 \\
Static-RQS & 12 & 12 & 12 & 12 & 0 & 12 \\
Static-Beta & 12 & 12 & 12 & 12 & 0 & 12 \\
RQS-GAS & 12 & 12 & 12 & 12 & 0 & 12 \\
RQS-Con & 12 & 12 & 12 & 12 & 0 & 12 \\
Beta-GAS & 12 & 12 & 12 & 12 & 0 & 12 \\
Beta-Con & 12 & 12 & 12 & 12 & 0 & 12 \\
\bottomrule
\end{tabular}
\caption{US inflation: number of the 12 LDF discount factors at which each test does not reject at 5\%.}
\label{tab:inf-tests-passed}
\end{table}

The results resemble the simulation scenarios with constant miscalibration. The raw pool is mainly shifted, a distortion that the two-parameter Beta map can represent, and the static Beta map, which estimates this correction from all past PITs, gives the largest gains in the log score, the CRPS and the left tail, together with the smallest KS distance (Table~\ref{tab:infl-scores}). The adaptive Beta maps come close in the log score, so allowing the map to vary over time adds little on average. The additional flexibility of the spline may be disadvantageous in this relatively short sample. With $3K+1=13$ parameters estimated from at most 150 PITs, the Static-RQS map reduces the mean log score by 0.045 relative to the raw pool, although the difference is not statistically significant. Even though the raw pool fails the uniformity tests, the relatively modest departure from uniformity suggests that the potential log-score gain may be limited. Where recalibration does make a significant difference is in the interval forecasts. The adaptive maps improve the 90\% and 95\% interval scores, with the largest gain of 0.125 in $\mathrm{IS}_{95}$ for Beta-GAS. These improvements suggest that time-varying recalibration can better balance interval width and coverage than a single static map.

\begin{table}[H]\centering\footnotesize
\begin{tabular}{lccccccc}\toprule
 Metric & Raw pool & Static-RQS & RQS-GAS & RQS-Con & Static-Beta & Beta-GAS & Beta-Con \\ \midrule
 KS & $0.132^{\dagger}$ & $0.056$ & $0.061$ & $0.062$ & $\mathbf{0.051}$ & $0.064$ & $0.062$ \\
 LogS & $-0.166$ & $-0.045$ & $+0.013$ & $+0.012$ & $\mathbf{+0.032}$ & $+0.026$ & $+0.029$ \\
 CRPS & $0.159$ & $+0.003$ & $+0.002$ & $+0.002$ & $\mathbf{+0.003}$ & $+0.000$ & $+0.001$ \\
 IS$_{90}$ & $1.233$ & $+0.018$ & $+0.027^{**}$ & $+0.033^{***}$ & $+0.014$ & $+0.056^{**}$ & $\mathbf{+0.058^{*}}$ \\
 IS$_{95}$ & $1.456$ & $+0.081$ & $+0.036^{***}$ & $+0.036^{**}$ & $+0.058$ & $\mathbf{+0.125^{***}}$ & $+0.091^{*}$ \\
 CSL$_{L}$ & $-0.118$ & $-0.014$ & $+0.012$ & $+0.016$ & $\mathbf{+0.044^{*}}$ & $+0.026^{*}$ & $+0.023$ \\
 CSL$_{R}$ & $-0.128$ & $-0.023$ & $-0.001$ & $-0.005$ & $-0.018$ & $-0.004$ & $\mathbf{+0.002}$ \\
\bottomrule\end{tabular}
\caption{US inflation, $\alpha=0.9$. KS: KS distance of the PIT from uniform ($^\dagger$ rejects at 5\%). Raw pool: level of each score; method columns: gain over the raw pool, positive is better. Stars: HAC Giacomini--White test at 10/5/1\%. Best method in bold.}
\label{tab:infl-scores}
\end{table}

On the shorter BPS evaluation window, BPS is the more accurate raw forecast: its mean log score exceeds that of the LDF pool by about 0.09 per quarter (Table~\ref{tab:infl-scores-combined}). Neither raw forecast is rejected by the KS test on this shorter window, providing limited evidence of departure from marginal uniformity. For BPS, the only map that improves the log score is RQS-Con, and every map worsens the CRPS. The static maps are the worst methods for both forecasts: the Static-RQS map lowers the log score of both, and the Static-Beta map that of BPS, significantly and by 0.07--0.13 per quarter. The adaptive maps still improve the 95\% interval score of the LDF pool, by 0.089 for Beta-GAS and 0.020 for RQS-GAS, both significantly.

\begin{table}[H]
\centering
\scriptsize
\begin{tabular}{llccccccc}
\toprule
Forecast & Metric & Raw pool & Static-RQS & RQS-GAS & RQS-Con & Static-Beta & Beta-GAS & Beta-Con \\ 
\midrule
LDF & KS & $0.118$ & $\mathbf{0.076}$ & $0.089$ & $0.077$ & $0.088$ & $0.086$ & $0.108$ \\
 & LogS & $-0.030$ & $-0.133^{***}$ & $+0.007$ & $-0.013$ & $-0.068^{*}$ & $\mathbf{+0.015}$ & $-0.012$ \\
 & CRPS & $0.140$ & $-0.004$ & $+0.000$ & $\mathbf{+0.001}$ & $-0.003$ & $-0.001$ & $-0.002$ \\
 & IS$_{90}$ & $1.018$ & $-0.074$ & $+0.017$ & $+0.012$ & $-0.092$ & $\mathbf{+0.032}$ & $+0.006$ \\
 & IS$_{95}$ & $1.173$ & $-0.037$ & $+0.020^{***}$ & $+0.009$ & $-0.107$ & $\mathbf{+0.089^{***}}$ & $+0.026$ \\
 & CSL$_{L}$ & $-0.083$ & $-0.058^{*}$ & $\mathbf{+0.016}$ & $-0.003$ & $-0.044$ & $+0.004$ & $-0.027$ \\
 & CSL$_{R}$ & $-0.111$ & $-0.028$ & $-0.006$ & $-0.004$ & $-0.024$ & $+0.000$ & $\mathbf{+0.003}$ \\
\midrule
BPS & KS & $0.119$ & $\mathbf{0.069}$ & $0.095$ & $0.086$ & $0.095$ & $0.107$ & $0.094$ \\
 & LogS & $0.064$ & $-0.107^{**}$ & $-0.002$ & $\mathbf{+0.015}$ & $-0.072^{**}$ & $-0.011$ & $-0.067$ \\
 & CRPS & $0.131$ & $-0.003^{*}$ & $\mathbf{-0.002}$ & $-0.002$ & $-0.003^{*}$ & $-0.002^{*}$ & $-0.004^{*}$ \\
 & IS$_{90}$ & $0.925$ & $-0.043^{*}$ & $-0.006$ & $\mathbf{-0.003}$ & $-0.058^{**}$ & $-0.006$ & $-0.043$ \\
 & IS$_{95}$ & $1.053$ & $-0.087^{*}$ & $+0.001$ & $-0.001$ & $-0.102^{*}$ & $\mathbf{+0.011}$ & $-0.044$ \\
 & CSL$_{L}$ & $-0.019$ & $-0.044$ & $+0.011$ & $\mathbf{+0.016}$ & $-0.046^{*}$ & $-0.007$ & $-0.062$ \\
 & CSL$_{R}$ & $-0.094$ & $-0.041^{*}$ & $-0.009$ & $-0.015$ & $-0.016$ & $\mathbf{-0.007}$ & $-0.008$ \\
\bottomrule
\end{tabular}
\caption{US inflation, LDF and BPS scores on the BPS evaluation window. Conventions as in Table~\ref{tab:infl-scores}.}
\label{tab:infl-scores-combined}
\end{table}

\section{Conclusion}
\label{sec:conclusion}

We have proposed a time-varying framework for recalibrating combined predictive distributions. A monotone map is applied to the combined PIT, leaving its weights unchanged, so the recalibration step can be placed on top of any combination. Since the recalibrated log score is the sum of the original log score and the log derivative of the map at the realised PIT, the map can be updated sequentially by score-driven filters, and the expected gain is bounded by the divergence of the PIT distribution from uniformity. Conditioning the map on the shape of the current forecast, through a neural network, allows it to anticipate changes in miscalibration rather than learn them from past PITs.

The simulations and applications suggest a simple guide to using the framework. The map family should match the shape of the miscalibration: the Beta map is sufficient for errors in location and spread, whereas multimodal miscalibration requires the spline. When the miscalibration is constant, a static map estimated on an expanding window is hard to beat, and the adaptive maps lose a little by fitting noise. When miscalibration changes with the weights of the pool, conditioning on the shape of the forecast can yield substantially larger gains than score-driven filtering alone. In the empirical applications, recalibration substantially improves PIT uniformity, particularly for the US inflation pool and the more miscalibrated BPS exchange-rate forecasts. These improvements are sometimes accompanied by statistically significant gains in predictive accuracy, although serial dependence may remain and recalibration does not improve every scoring rule.

In practice, recalibration can be implemented as a post-processing step without re-estimating the underlying combination model. It is most useful when the raw forecast is clearly miscalibrated, or when its miscalibration changes over time. When the forecast is already close to calibrated, there is little to gain, and the more flexible maps can lose slightly because their parameters are estimated with noise.

Our results also point to three directions for further work. First, the maps do not explicitly target serial dependence in the PITs, and the adaptive updating may introduce additional dependence. Including lagged PITs among the features of the inner map would allow the correction to respond to recent forecast errors. Second, the spline may overfit in short samples; shrinking its bins towards equal widths with a penalty that weakens as the sample grows, or selecting the number of bins by dynamic model averaging, would let its flexibility grow with the data. Third, the framework is designed for one-step-ahead forecasts of a single variable. Multi-step forecasts could be handled by updating the filters with a delay, and multivariate forecasts by recalibrating each marginal and correcting the dependence separately, for example through a copula.

\section*{Data and code availability}
The data and the code reproducing the results in this paper are freely available on \textcolor{Blue}{\href{https://github.com/callum-0214/recalibration}{https://github.com/callum-0214/recalibration}}.

\section*{Funding}
This work was supported by the Martingale Foundation through a Martingale PhD Scholarship awarded to Callum Wu.

\newpage
\appendix
\section{Expected log-score gain}
\label{app:expectedlog}
Assume that $f_t > 0$ wherever $g_t > 0$, that $\tau'_{\theta_t} > 0$ on $(0,1)$, and that the divergences below are finite. Expectations $\mathbb{E}_{t-1}$ are taken under $G_t$. Since $F_t$ is strictly increasing, 
\begin{equation*}
\Pi_t(u) = \Pr_{t-1}(y_t \le F_t^{-1}(u)) = G_t(F_t^{-1}(u)),
\end{equation*}
and differentiating gives
\begin{equation*}
      \pi_t(u) = \frac{g_t\bigl(F_t^{-1}(u)\bigr)}{f_t\bigl(F_t^{-1}(u)\bigr)}.
\end{equation*}
Evaluating at $u = u_t = F_t(y_t)$ gives $\log \pi_t(u_t) = \log g_t(y_t) - \log f_t(y_t)$. The Kullback--Leibler divergence between $G_t$ and $F_t$ is
\begin{equation*}
  \mathrm{KL}(G_t \,\|\, F_t) = \int g_t \log \frac{g_t}{f_t}\, \mathrm{d}y = \int_0^1 \pi_t \log \pi_t\, \mathrm{d}u = \mathrm{KL}(\Pi_t \,\|\, \mathrm{U}).    
\end{equation*}
The recalibrated density is $\tilde f_t(y) = \tau'_{\theta_t}\bigl(F_t(y)\bigr) f_t(y)$, so 
\begin{equation*}
  \mathrm{KL}\bigl(G_t \,\|\, \tilde F_t\bigr)
  = \int g_t(y) \log \frac{g_t(y)}{\tau'_{\theta_t}\bigl(F_t(y)\bigr) f_t(y)}\, \mathrm{d}y
  = \int_0^1 \pi_t(u) \log \frac{\pi_t(u)}{\tau'_{\theta_t}(u)}\, \mathrm{d}u
  = \mathrm{KL}\bigl(\Pi_t \,\|\, \tau_{\theta_t}\bigr),
\end{equation*}
the divergence of $\Pi_t$ from the distribution on $(0,1)$ with CDF $\tau_{\theta_t}$ and density $\tau'_{\theta_t}$.

Taking expectations, and adding and subtracting $\log g_t(y_t)$,
\begin{equation*}
  \mathbb{E}_{t-1}[\ell_t]
  = \mathbb{E}_{t-1}\Bigl[\log \frac{g_t(y_t)}{f_t(y_t)}\Bigr] - \mathbb{E}_{t-1}\Bigl[\log \frac{g_t(y_t)}{\tilde f_t(y_t)}\Bigr]
  = \mathrm{KL}(\Pi_t \,\|\, \mathrm{U}) - \mathrm{KL}\bigl(\Pi_t \,\|\, \tau_{\theta_t}\bigr).
\end{equation*}

\newpage

\section{Score of the Beta map}
\label{app:beta_score}

This appendix derives the score \eqref{eq:beta_score} of the Beta map. To lighten notation we drop the subscript $t$ and write $u$ for the PIT, $(a, b)$ for the shape parameters and $\theta = (\theta_a, \theta_b)'$ for the unconstrained parameter.

The log-score gain is the log of the Beta density at $u$,
\begin{equation}
    \ell(a, b) = (a - 1)\log u + (b - 1)\log(1 - u) - \log B(a, b),
  \label{eq:app_beta_ll}
\end{equation}
with $\log B(a, b) = \log\Gamma(a) + \log\Gamma(b) - \log\Gamma(a + b)$. Let $\psi(x) = \mathrm{d}\log\Gamma(x)/\mathrm{d}x$ denote the digamma function. Differentiating $\log B(a,b)$ term by term gives
\[
  \frac{\partial \log B(a,b)}{\partial a} = \psi(a) - \psi(a+b),
  \qquad
  \frac{\partial \log B(a,b)}{\partial b} = \psi(b) - \psi(a+b),
\]
where the term $\psi(a+b)$ comes from the chain rule applied to $\log\Gamma(a+b)$. The first two terms of \eqref{eq:app_beta_ll} are linear in $a$ and $b$, so
\begin{equation}
  \frac{\partial \ell}{\partial a} = \log u - \psi(a) + \psi(a + b),
  \qquad
  \frac{\partial \ell}{\partial b} = \log(1 - u) - \psi(b) + \psi(a + b).
  \label{eq:app_beta_dab}
\end{equation}
By \eqref{eq:beta_link}, $a = c + \mathrm{softplus}(\theta_a + s)$ and $b = c + \mathrm{softplus}(\theta_b + s)$. For $\mathrm{softplus}(x) = \log(1 + e^x)$,
\[
  \frac{\mathrm{d}}{\mathrm{d}x}\,\mathrm{softplus}(x) = \frac{e^x}{1 + e^x} = \frac{1}{1 + e^{-x}} = \Lambda(x),
\]
the logistic function. Since $c$ and $s$ are constants,
\begin{equation}
  \frac{\partial a}{\partial \theta_a} = \Lambda(\theta_a + s),
  \qquad
  \frac{\partial b}{\partial \theta_b} = \Lambda(\theta_b + s),
  \qquad
  \frac{\partial a}{\partial \theta_b} = \frac{\partial b}{\partial \theta_a} = 0.
  \label{eq:app_beta_link}
\end{equation}
The chain rule combines \eqref{eq:app_beta_dab} and \eqref{eq:app_beta_link} into
\begin{equation*}
  \nabla = \frac{\partial \ell}{\partial \theta} =
  \begin{pmatrix}
    \dfrac{\partial \ell}{\partial a}\,\dfrac{\partial a}{\partial \theta_a} \\[10pt]
    \dfrac{\partial \ell}{\partial b}\,\dfrac{\partial b}{\partial \theta_b}
  \end{pmatrix}
  =
  \begin{pmatrix}
    \bigl[\log u - \psi(a) + \psi(a + b)\bigr]\,\Lambda(\theta_a + s) \\[4pt]
    \bigl[\log(1 - u) - \psi(b) + \psi(a + b)\bigr]\,\Lambda(\theta_b + s)
  \end{pmatrix}.
\end{equation*}

\newpage
\section{Score of the rational quadratic spline}
\label{app:rqs_score}
The log-score gain is formally written as
\begin{equation}
    \ell_t =\log\frac{\big(s^{(k)}_t\big)^2\Big[\delta^{(k+1)}_t\xi_t^2 + 2s^{(k)}_t\xi_t(1-\xi_t) + \delta^{(k)}_t(1-\xi_t)^2 \Big]}{\bigl(s^{(k)}_t + (\delta^{(k+1)}_t + \delta^{(k)}_t - 2s^{(k)}_t)\xi_t(1-\xi_t)\bigr)^2},
    \label{eq:rqs_ll}
\end{equation}
with
\[
\xi_t = \frac{u_t - u^{(k)}_t}{u^{(k+1)}_t - u^{(k)}_t}
 \quad \text{and} \quad s^{(k)}_t = \frac{v^{(k+1)}_t - v^{(k)}_t}{u^{(k+1)}_t - u^{(k)}_t}.
\]
Fix $t$ and let $u_t \in [u^{(k)}_t, u^{(k+1)}_t)$. To lighten notation we drop the subscript $t$ and write
\begin{equation*}
\footnotesize
  W = u^{(k+1)}-u^{(k)}, \quad H = v^{(k+1)}-v^{(k)}, \quad U = u^{(k)}, \quad
  s = \frac{H}{W}, \quad \xi = \frac{u - U}{W}, \quad d_0 = \delta^{(k)}, \quad d_1 = \delta^{(k+1)}.    
\end{equation*}

\paragraph{Step 1: derivatives with respect to $(\xi, s, d_0, d_1)$}
Write \eqref{eq:rqs_ll} as $\ell = 2\log s + \log N - 2\log D$, where
\begin{align*}
  N &= d_1\xi^2 + 2s\,\xi(1-\xi) + d_0(1-\xi)^2, \\[5pt]
  D &= s + (d_0 + d_1 - 2s)\,\xi(1-\xi).
\end{align*}
Differentiating term by term,
\begin{align}
  \frac{\partial \ell}{\partial s} &= \frac{2}{s} + \frac{2\xi(1-\xi)}{N} - \frac{2\bigl[1 - 2\xi(1-\xi)\bigr]}{D}, \label{eq:rqs_ds}\\
  \frac{\partial \ell}{\partial d_0} &= \frac{(1-\xi)^2}{N} - \frac{2\xi(1-\xi)}{D}, \label{eq:rqs_dd0}\\
  \frac{\partial \ell}{\partial d_1} &= \frac{\xi^2}{N} - \frac{2\xi(1-\xi)}{D}, \label{eq:rqs_dd1}\\
  \frac{\partial \ell}{\partial \xi} &= \frac{2\bigl[d_1\xi + s(1-2\xi) - d_0(1-\xi)\bigr]}{N} - \frac{2(d_0 + d_1 - 2s)(1-2\xi)}{D}. \label{eq:rqs_dxi}
\end{align}

\paragraph{Step 2: derivatives with respect to widths, heights and knot derivatives}
Since $\xi = (u - U)/W$ and $s = H/W$,
\[
  \frac{\partial \xi}{\partial U} = -\frac{1}{W}, \qquad
  \frac{\partial \xi}{\partial W} = -\frac{\xi}{W}, \qquad
  \frac{\partial s}{\partial W} = -\frac{s}{W}, \qquad
  \frac{\partial s}{\partial H} = \frac{1}{W}.
\]
The left knot $U = \sum_{j \le k}w_j$ depends on the widths of all bins to the left of $u$, while $W$ is the width of the current bin and the output knot $v^{(k)}$ does not enter $\ell$. Hence, for $j = 1, \dots, K$,
\begin{equation}
  g^{u}_j \equiv \frac{\partial \ell}{\partial w_j} =
  \begin{cases}
    -\dfrac{1}{W}\dfrac{\partial \ell}{\partial \xi}, & j \le k,\\[8pt]
    -\dfrac{1}{W}\Bigl[\xi\,\dfrac{\partial \ell}{\partial \xi} + s\,\dfrac{\partial \ell}{\partial s}\Bigr], & j = k+1,\\[8pt]
    0, & j > k+1,
  \end{cases}
  \qquad
  g^{v}_j \equiv \frac{\partial \ell}{\partial h_j} =
  \begin{cases}
    \dfrac{1}{W}\dfrac{\partial \ell}{\partial s}, & j = k+1,\\[8pt]
    0, & \text{otherwise},
  \end{cases}
  \label{eq:rqs_gwh}
\end{equation}
and only the two knot derivatives of the current bin enter $\ell$, with derivatives \eqref{eq:rqs_dd0} and \eqref{eq:rqs_dd1}.

\paragraph{Step 3: derivatives with respect to $\theta$}
Let $p^{u} = \mathrm{softmax}(\theta^{u})$ and $p^{v} = \mathrm{softmax}(\theta^{v})$. From \eqref{eq:rqs_link},
$\partial w_j/\partial\theta^{u}_i = \,p^{u}_j(\mathbbm{1}\{i = j\} - p^{u}_i)$, and similarly for the heights. Summing over $j$,
\begin{equation}
  \frac{\partial \ell}{\partial \theta^{u}_i} = \,p^{u}_i\Bigl(g^{u}_i - \sum_{j=1}^{K} p^{u}_j g^{u}_j\Bigr),
  \qquad
  \frac{\partial \ell}{\partial \theta^{v}_i} = \,p^{v}_i\Bigl(g^{v}_i - \sum_{j=1}^{K} p^{v}_j g^{v}_j\Bigr),
  \label{eq:rqs_dtheta_wh}
\end{equation}
for $i = 1, \dots, K$. Each component compares the gradient for bin $i$ with its softmax-weighted average, so every component of $\theta^{u}$ and $\theta^{v}$ is updated even though most $g^{u}_j$ and $g^{v}_j$ are zero. Finally, since $\mathrm{d}\,\mathrm{softplus}_{\varrho}(x)/\mathrm{d}x = \Lambda(\varrho x)$, with $\Lambda$ the logistic function,
\begin{equation}
  \frac{\partial \ell}{\partial \theta^{d}_i} =
  \begin{cases}
    \dfrac{\partial \ell}{\partial d_0}\,\Lambda(\varrho\theta^{d}_k), & i = k,\\[8pt]
    \dfrac{\partial \ell}{\partial d_1}\,\Lambda(\varrho\theta^{d}_{k+1}), & i = k+1,\\[8pt]
    0, & \text{otherwise},
  \end{cases}
  \qquad i = 0, \dots, K.
  \label{eq:rqs_dtheta_d}
\end{equation}
Stacking \eqref{eq:rqs_dtheta_wh} and \eqref{eq:rqs_dtheta_d} gives the score $\nabla_t \in \mathbb{R}^{3K+1}$.

\newpage

\section{Simulation results}
\label{app:simulation}

\begin{figure}[H]
    \centering
    \includegraphics[width=\linewidth]{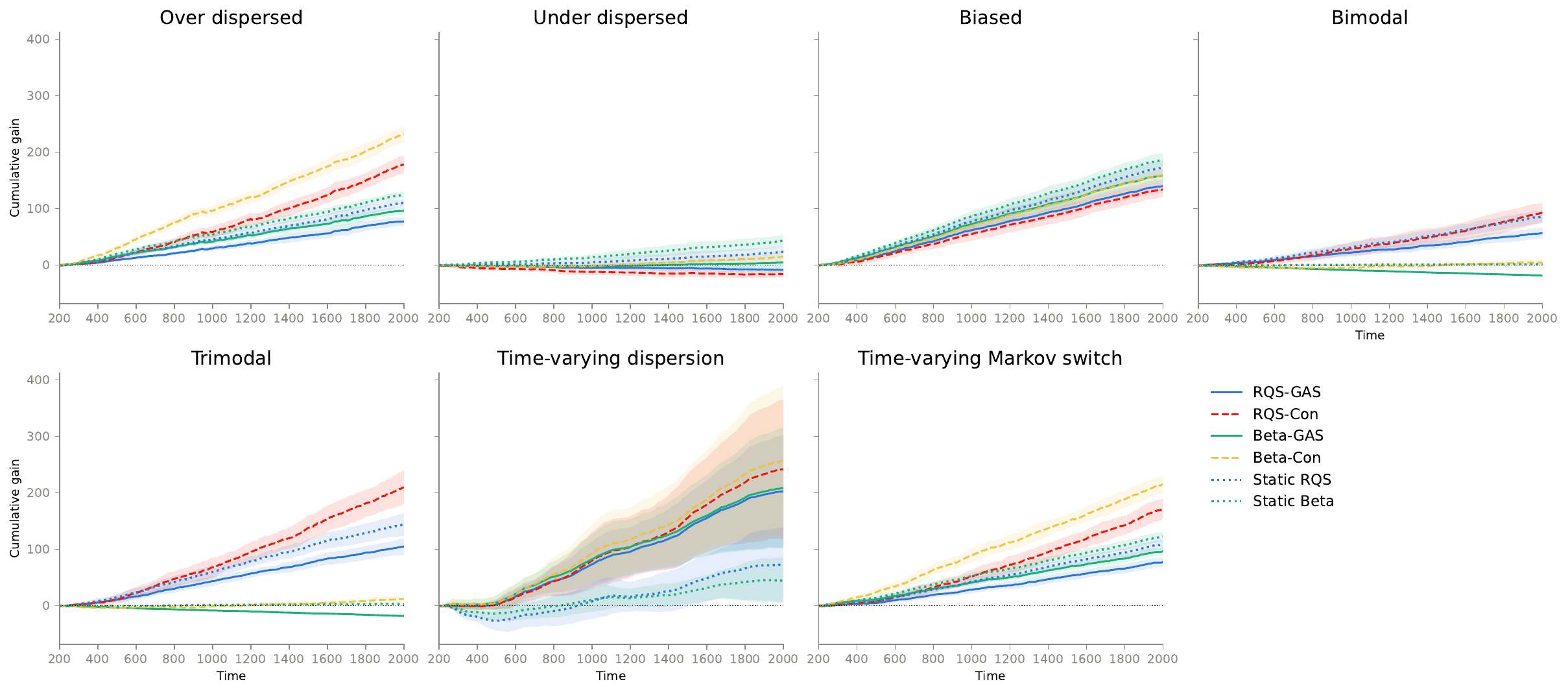}
    \caption{Cumulative log-score gain over the raw pool, mean over ten replications with 95\% bands across replications, $\alpha = 0.9$. Solid: GAS; dashed: convolution; dotted: static map of the same family.}
    \label{fig:cum-gain}
\end{figure}

\begin{figure}[H]
    \centering
    \includegraphics[width=\linewidth]{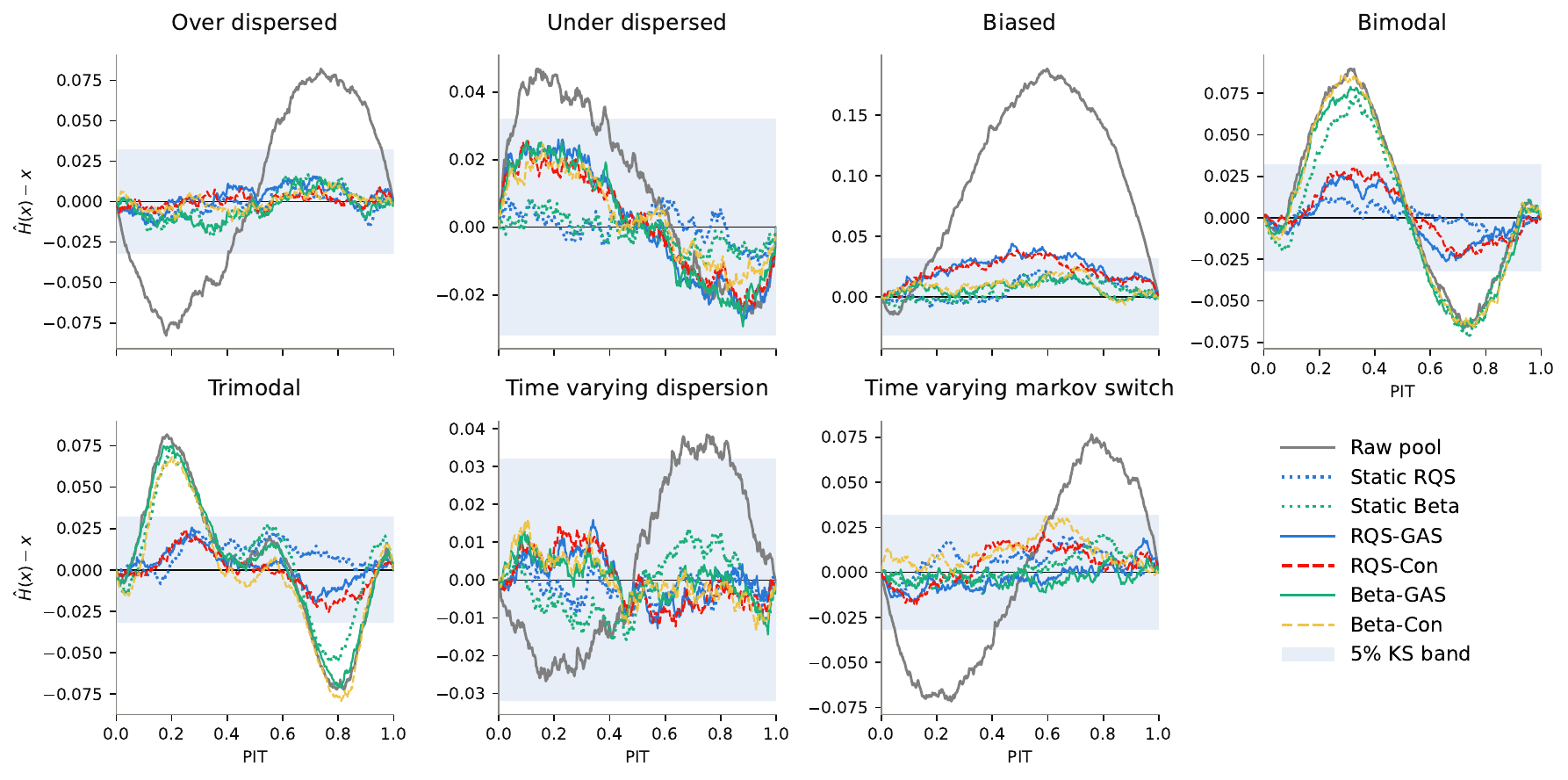}
    \caption{Deviation of the empirical CDF of the PITs from the uniform CDF for the raw pool and the six methods; the shaded band is the 5\% Kolmogorov–Smirnov acceptance region.}
    \label{fig:sim-ecdf}
\end{figure}

\begin{table}[H]\centering\tiny
\begin{tabular}{llccccccc}\toprule
Scenario & Metric & Raw pool & Static-RQS & RQS-GAS & RQS-Con & Static-Beta & Beta-GAS & Beta-Con \\ \midrule
Over-dispersed & KS & $0.082$ [1.0] & $0.019$ [0.0] & $0.018$ [0.0] & $\mathbf{0.015}$ [0.0] & $0.023$ [0.1] & $0.016$ [0.0] & $0.020$ [0.1] \\
 & LogS & $-1.1912$ & $+0.0612^{***}$ & $+0.0429^{***}$ & $+0.0989^{***}$ & $+0.0691^{***}$ & $+0.0534^{***}$ & $\mathbf{+0.1285^{***}}$ \\
 & CRPS & $0.4059$ & $+0.0100^{***}$ & $+0.0057^{***}$ & $+0.0229^{***}$ & $+0.0100^{***}$ & $+0.0056^{***}$ & $\mathbf{+0.0308^{***}}$ \\
 & IS$_{90}$ & $3.2949$ & $+0.3299^{***}$ & $+0.2619^{***}$ & $+0.3605^{***}$ & $+0.3295^{***}$ & $+0.2777^{***}$ & $\mathbf{+0.3910^{***}}$ \\
 & IS$_{95}$ & $3.8463$ & $+0.4298^{***}$ & $+0.3298^{***}$ & $+0.4117^{***}$ & $+0.4426^{***}$ & $+0.3748^{***}$ & $\mathbf{+0.4629^{***}}$ \\
 & CSL$_{L}$ & $-0.2807$ & $+0.0187^{***}$ & $+0.0139^{***}$ & $+0.0315^{***}$ & $+0.0207^{***}$ & $+0.0168^{***}$ & $\mathbf{+0.0381^{***}}$ \\
 & CSL$_{R}$ & $-0.2665$ & $+0.0232^{***}$ & $+0.0175^{***}$ & $+0.0308^{***}$ & $+0.0246^{***}$ & $+0.0196^{***}$ & $\mathbf{+0.0391^{***}}$ \\
\addlinespace
Under-dispersed & KS & $0.047$ [1.0] & $0.022$ [0.0] & $0.028$ [0.2] & $0.027$ [0.1] & $\mathbf{0.020}$ [0.1] & $0.032$ [0.4] & $0.026$ [0.1] \\
 & LogS & $-1.1781$ & $+0.0130^{***}$ & $-0.0047^{***}$ & $-0.0090^{***}$ & $\mathbf{+0.0239^{***}}$ & $+0.0026$ & $+0.0082^{***}$ \\
 & CRPS & $0.4305$ & $+0.0015^{***}$ & $-0.0009^{***}$ & $-0.0010^{***}$ & $\mathbf{+0.0016^{***}}$ & $-0.0028^{***}$ & $-0.0027^{***}$ \\
 & IS$_{90}$ & $3.2247$ & $+0.0706^{***}$ & $+0.0202^{**}$ & $+0.0272^{***}$ & $\mathbf{+0.0719^{***}}$ & $-0.0001$ & $+0.0133$ \\
 & IS$_{95}$ & $3.7369$ & $+0.1245^{***}$ & $+0.0307^{**}$ & $+0.0367^{***}$ & $\mathbf{+0.1359^{***}}$ & $+0.0158$ & $+0.0456^{**}$ \\
 & CSL$_{L}$ & $-0.2563$ & $+0.0044^{***}$ & $-0.0002$ & $-0.0001$ & $\mathbf{+0.0065^{***}}$ & $+0.0006$ & $+0.0017^{**}$ \\
 & CSL$_{R}$ & $-0.2407$ & $+0.0027^{***}$ & $-0.0007$ & $-0.0014^{***}$ & $\mathbf{+0.0044^{***}}$ & $-0.0002$ & $+0.0006$ \\
\addlinespace
Biased & KS & $0.173$ [1.0] & $0.026$ [0.2] & $0.039$ [0.8] & $0.034$ [0.8] & $0.029$ [0.3] & $\mathbf{0.023}$ [0.1] & $0.027$ [0.1] \\
 & LogS & $-0.8822$ & $+0.0959^{***}$ & $+0.0777^{***}$ & $+0.0743^{***}$ & $\mathbf{+0.1035^{***}}$ & $+0.0880^{***}$ & $+0.0883^{***}$ \\
 & CRPS & $0.3203$ & $+0.0251^{***}$ & $+0.0216^{***}$ & $+0.0221^{***}$ & $\mathbf{+0.0252^{***}}$ & $+0.0220^{***}$ & $+0.0220^{***}$ \\
 & IS$_{90}$ & $2.4510$ & $\mathbf{+0.2790^{***}}$ & $+0.2192^{***}$ & $+0.2284^{***}$ & $+0.2777^{***}$ & $+0.2368^{***}$ & $+0.2332^{***}$ \\
 & IS$_{95}$ & $2.8248$ & $+0.3368^{***}$ & $+0.2454^{***}$ & $+0.2503^{***}$ & $\mathbf{+0.3403^{***}}$ & $+0.2756^{***}$ & $+0.2759^{***}$ \\
 & CSL$_{L}$ & $-0.1881$ & $+0.0089^{***}$ & $+0.0046^{***}$ & $+0.0056^{***}$ & $\mathbf{+0.0095^{***}}$ & $+0.0064^{***}$ & $+0.0067^{***}$ \\
 & CSL$_{R}$ & $-0.2076$ & $+0.0398^{***}$ & $+0.0355^{***}$ & $+0.0350^{***}$ & $\mathbf{+0.0412^{***}}$ & $+0.0385^{***}$ & $+0.0386^{***}$ \\
\addlinespace
Bimodal & KS & $0.080$ [1.0] & $\mathbf{0.021}$ [0.1] & $0.031$ [0.4] & $0.028$ [0.0] & $0.071$ [1.0] & $0.075$ [1.0] & $0.077$ [1.0] \\
 & LogS & $-1.6196$ & $+0.0477^{***}$ & $+0.0315^{***}$ & $\mathbf{+0.0513^{***}}$ & $+0.0009^{**}$ & $-0.0104^{***}$ & $+0.0024$ \\
 & CRPS & $0.6837$ & $+0.0046^{***}$ & $+0.0010^{*}$ & $\mathbf{+0.0055^{***}}$ & $+0.0007^{***}$ & $-0.0057^{***}$ & $-0.0040^{***}$ \\
 & IS$_{90}$ & $4.3567$ & $+0.0135^{***}$ & $-0.0157^{***}$ & $\mathbf{+0.0566^{***}}$ & $-0.0379^{***}$ & $-0.0460^{***}$ & $+0.0115^{*}$ \\
 & IS$_{95}$ & $4.8470$ & $+0.0413^{***}$ & $+0.0267^{**}$ & $\mathbf{+0.0829^{***}}$ & $-0.0590^{***}$ & $-0.0372^{***}$ & $+0.0068$ \\
 & CSL$_{L}$ & $-0.3048$ & $+0.0022^{*}$ & $+0.0001$ & $\mathbf{+0.0029^{**}}$ & $-0.0020^{***}$ & $-0.0028^{***}$ & $-0.0013^{**}$ \\
 & CSL$_{R}$ & $-0.2992$ & $+0.0039^{***}$ & $+0.0024^{***}$ & $\mathbf{+0.0084^{***}}$ & $-0.0020^{***}$ & $-0.0023^{***}$ & $-0.0014$ \\
\addlinespace
Trimodal & KS & $0.083$ [1.0] & $\mathbf{0.023}$ [0.0] & $0.026$ [0.1] & $0.028$ [0.2] & $0.072$ [1.0] & $0.077$ [1.0] & $0.081$ [1.0] \\
 & LogS & $-1.7138$ & $+0.0800^{***}$ & $+0.0582^{***}$ & $\mathbf{+0.1165^{***}}$ & $+0.0019^{***}$ & $-0.0100^{***}$ & $+0.0067^{***}$ \\
 & CRPS & $0.7538$ & $+0.0046^{***}$ & $+0.0008$ & $\mathbf{+0.0066^{***}}$ & $+0.0012^{***}$ & $-0.0056^{***}$ & $-0.0023^{**}$ \\
 & IS$_{90}$ & $4.5799$ & $+0.0074^{*}$ & $-0.0309^{***}$ & $\mathbf{+0.1417^{***}}$ & $-0.0681^{***}$ & $-0.0751^{***}$ & $+0.0107$ \\
 & IS$_{95}$ & $5.0572$ & $+0.0906^{***}$ & $+0.0410^{***}$ & $\mathbf{+0.2100^{***}}$ & $-0.1180^{***}$ & $-0.0696^{***}$ & $-0.0206^{*}$ \\
 & CSL$_{L}$ & $-0.3059$ & $+0.0054^{***}$ & $+0.0018^{***}$ & $\mathbf{+0.0114^{***}}$ & $-0.0039^{***}$ & $-0.0038^{***}$ & $-0.0015^{**}$ \\
 & CSL$_{R}$ & $-0.3018$ & $+0.0093^{***}$ & $+0.0050^{***}$ & $\mathbf{+0.0191^{***}}$ & $-0.0026^{***}$ & $-0.0039^{***}$ & $-0.0017^{*}$ \\
\addlinespace
Time-varying & KS & $0.082$ [1.0] & $0.037$ [0.7] & $0.029$ [0.5] & $\mathbf{0.023}$ [0.1] & $0.051$ [0.7] & $0.024$ [0.3] & $0.025$ [0.3] \\
  dispersion & LogS & $-1.1518$ & $+0.0401^{**}$ & $+0.1125^{***}$ & $+0.1343^{***}$ & $+0.0247^{**}$ & $+0.1156^{***}$ & $\mathbf{+0.1426^{***}}$ \\
 & CRPS & $0.4290$ & $+0.0336^{***}$ & $+0.0476^{***}$ & $+0.0567^{***}$ & $+0.0237^{***}$ & $+0.0473^{***}$ & $\mathbf{+0.0593^{***}}$ \\
 & IS$_{90}$ & $4.0360$ & $+0.3181^{***}$ & $+0.9256^{***}$ & $+0.9972^{***}$ & $+0.3770^{***}$ & $+0.9992^{***}$ & $\mathbf{+1.0905^{***}}$ \\
 & IS$_{95}$ & $4.6458$ & $+0.2247^{***}$ & $+0.9310^{***}$ & $+0.9874^{***}$ & $+0.3603^{***}$ & $+1.1113^{***}$ & $\mathbf{+1.2149^{***}}$ \\
 & CSL$_{L}$ & $-0.2876$ & $+0.0187^{***}$ & $+0.0382^{***}$ & $+0.0437^{***}$ & $+0.0110^{**}$ & $+0.0372^{***}$ & $\mathbf{+0.0459^{***}}$ \\
 & CSL$_{R}$ & $-0.2800$ & $+0.0218^{***}$ & $+0.0378^{***}$ & $+0.0477^{***}$ & $+0.0133^{***}$ & $+0.0403^{***}$ & $\mathbf{+0.0506^{***}}$ \\
\addlinespace
Time-varying  & KS & $0.083$ [1.0] & $0.022$ [0.0] & $0.018$ [0.0] & $0.017$ [0.0] & $0.024$ [0.0] & $\mathbf{0.015}$ [0.0] & $0.023$ [0.1] \\
Markov switch & LogS & $-1.2064$ & $+0.0604^{***}$ & $+0.0431^{***}$ & $+0.0954^{***}$ & $+0.0680^{***}$ & $+0.0536^{***}$ & $\mathbf{+0.1197^{***}}$ \\
 & CRPS & $0.4124$ & $+0.0103^{***}$ & $+0.0058^{***}$ & $+0.0224^{***}$ & $+0.0103^{***}$ & $+0.0060^{***}$ & $\mathbf{+0.0300^{***}}$ \\
 & IS$_{90}$ & $3.3365$ & $+0.3221^{***}$ & $+0.2654^{***}$ & $+0.3470^{***}$ & $+0.3248^{***}$ & $+0.2814^{***}$ & $\mathbf{+0.3562^{***}}$ \\
 & IS$_{95}$ & $3.8950$ & $+0.4256^{***}$ & $+0.3372^{***}$ & $+0.4123^{***}$ & $\mathbf{+0.4383^{***}}$ & $+0.3821^{***}$ & $+0.4192^{***}$ \\
 & CSL$_{L}$ & $-0.2802$ & $+0.0195^{***}$ & $+0.0143^{***}$ & $+0.0281^{***}$ & $+0.0206^{***}$ & $+0.0179^{***}$ & $\mathbf{+0.0355^{***}}$ \\
 & CSL$_{R}$ & $-0.2780$ & $+0.0217^{***}$ & $+0.0159^{***}$ & $+0.0287^{***}$ & $+0.0225^{***}$ & $+0.0179^{***}$ & $\mathbf{+0.0382^{***}}$ \\
\bottomrule\end{tabular}
\caption{Simulation results at $\alpha=0.9$, averaged over 10 replications. KS: mean Kolmogorov--Smirnov distance of the PIT from uniform, with the share of replications rejecting at 5\% in brackets. Raw pool: mean level of each score. Method columns: mean gain over the raw pool, oriented so that positive is better; stars from the $t$-test of the gain across replications at 10/5/1\%. Best method in bold. Static-RQS, Static-Beta: static maps of each family.}
\label{tab:sim_scores_appendix}
\end{table}

\newpage

\section{Bayesian predictive synthesis}
\label{app:bps}
\paragraph{Settings}
Table~\ref{tab:bps-settings} lists the settings used in each application. The discount factors and the prior on the synthesis coefficients follow the authors' code. The prior on the observation variance, $s_0$, is set on the scale of each series. Each run uses 8000 Monte Carlo draws, of which the first 3000 are discarded as burn-in, and is repeated at every forecast origin on an expanding
window.

\begin{table}[H]
\centering
\small
\begin{tabular}{l c c}
\toprule
 & Exchange rates & US inflation \\
\midrule
Agents $J$                         & 32 & 4  \\
State discount $\delta$            & 0.99 & 0.95 \\
Volatility discount $\beta$        & 0.99 & 0.99 \\
Prior mean $m_0$                   & $(0, 1/32, \dots, 1/32)'$ & $(0, 1/4, \dots, 1/4)'$ \\
Prior scale $C_0$                  & $0.297\, I_{33}$ & $I_5$ \\
Prior d.o.f.\ $n_0 = 1/(1-\beta)$  & 100 & 100 \\
Prior obs.\ variance $s_0$         & 0.0014 & 0.01 \\
Training window                    & 50  & 50  \\
Evaluation window                  & 312  & 100 \\
MCMC Iterations                   & 8000 & 8000\\
MCMC burn-in                      & 3000  & 3000 \\
\bottomrule
\end{tabular}
\caption{Dynamic BPS settings. $\delta$ discounts the state (synthesis
coefficients) and $\beta$ the observation variance. For exchange rates the Gaussian
agent densities are represented as Student-$t$ with 1000 degrees of freedom,
and $s_0$ is the sample variance of the 50 training months.}
\label{tab:bps-settings}
\end{table}

\vspace{1cm}

\begin{table}[H]
\centering
\footnotesize
\begin{tabular}{llcccccc}
\toprule
Currency & Raw pool & RMSE & mean LogS & KS & KS $p$ & 90\% coverage & 90\% width \\
\midrule
\multirow{2}{*}{AUD} & LDF & $\mathbf{0.0361}$ & $\mathbf{1.9238}$ & $\mathbf{0.0573}$ & $\mathbf{0.4437}$ & $\mathbf{0.9144}$ & $\mathbf{0.1204}$ \\
 & BPS & $0.0362$ & $1.9137$ & $0.0834$ & $0.0864$ & $0.9234$ & $0.1307$ \\
\addlinespace
\multirow{2}{*}{CAD} & LDF & $0.0272$ & $2.1748$ & $\mathbf{0.0524}$ & $\mathbf{0.5566}$ & $\mathbf{0.8829}$ & $\mathbf{0.0860}$ \\
 & BPS & $\mathbf{0.0270}$ & $\mathbf{2.1847}$ & $0.0790$ & $0.1182$ & $0.9234$ & $0.0967$ \\
\addlinespace
\multirow{2}{*}{EUR} & LDF & $0.0287$ & $\mathbf{2.1703}$ & $\mathbf{0.0838}$ & $\mathbf{0.0838}$ & $\mathbf{0.9189}$ & $\mathbf{0.0995}$ \\
 & BPS & $\mathbf{0.0281}$ & $2.1382$ & $0.1081$ & $0.0103$ & $0.9369$ & $0.1133$ \\
\addlinespace
\multirow{2}{*}{JPY} & LDF & $0.0265$ & $\mathbf{2.1984}$ & $\mathbf{0.0826}$ & $\mathbf{0.0911}$ & $\mathbf{0.9324}$ & $\mathbf{0.1044}$ \\
 & BPS & $\mathbf{0.0259}$ & $2.1799$ & $0.0879$ & $0.0608$ & $0.9505$ & $0.1100$ \\
\addlinespace
\multirow{2}{*}{NZD} & LDF & $0.0381$ & $\mathbf{1.8697}$ & $\mathbf{0.0646}$ & $\mathbf{0.3002}$ & $\mathbf{0.9144}$ & $\mathbf{0.1263}$ \\
 & BPS & $\mathbf{0.0378}$ & $1.8610$ & $0.0673$ & $0.2551$ & $0.9369$ & $0.1380$ \\
\addlinespace
\multirow{2}{*}{NOK} & LDF & $0.0346$ & $1.9289$ & $\mathbf{0.0365}$ & $\mathbf{0.9185}$ & $0.8739$ & $\mathbf{0.1110}$ \\
 & BPS & $\mathbf{0.0340}$ & $\mathbf{1.9447}$ & $0.0660$ & $0.2766$ & $\mathbf{0.9189}$ & $0.1243$ \\
\addlinespace
\multirow{2}{*}{SEK} & LDF & $0.0332$ & $\mathbf{1.9992}$ & $\mathbf{0.0547}$ & $\mathbf{0.5015}$ & $\mathbf{0.9144}$ & $\mathbf{0.1114}$ \\
 & BPS & $\mathbf{0.0330}$ & $1.9755$ & $0.0690$ & $0.2309$ & $0.9414$ & $0.1292$ \\
\addlinespace
\multirow{2}{*}{CHF} & LDF & $0.0307$ & $\mathbf{2.0995}$ & $\mathbf{0.0873}$ & $\mathbf{0.0641}$ & $\mathbf{0.9279}$ & $\mathbf{0.1065}$ \\
 & BPS & $\mathbf{0.0297}$ & $2.0844$ & $0.1029$ & $0.0169$ & $0.9640$ & $0.1202$ \\
\addlinespace
\multirow{2}{*}{GBP} & LDF & $\mathbf{0.0254}$ & $\mathbf{2.2491}$ & $\mathbf{0.0616}$ & $\mathbf{0.3543}$ & $\mathbf{0.9279}$ & $\mathbf{0.0874}$ \\
 & BPS & $\mathbf{0.0254}$ & $2.2201$ & $0.0839$ & $0.0830$ & $0.9595$ & $0.1008$ \\
\bottomrule
\end{tabular}
\caption{Exchange rates: raw LDF pool and raw BPS on the BPS evaluation window. Lower is better for RMSE, KS distance and 90\% interval width; higher is better for the mean log score and the KS $p$-value; coverage closer to 0.90 is better. The better of the two forecasts for each metric is in bold.}
\label{tab:raw-pool-comparison}
\end{table}

\begin{table}[H]\centering\tiny
\begin{tabular}{llccccccc}\toprule
Currency & Metric & Raw pool & Static-RQS & RQS-GAS & RQS-Con & Static-Beta & Beta-GAS & Beta-Con \\ \midrule
AUD & KS & $0.083$ & $0.046$ & $\mathbf{0.036}$ & $0.037$ & $0.059$ & $0.046$ & $0.047$ \\
 & LogS & $1.9137$ & $-0.0022$ & $+0.0018$ & $+0.0022$ & $-0.0158$ & $\mathbf{+0.0130}$ & $+0.0129$ \\
 & CRPS & $0.0198$ & $-0.0000$ & $\mathbf{+0.0001}$ & $+0.0001$ & $+0.0000$ & $+0.0001$ & $+0.0001$ \\
 & IS$_{90}$ & $0.1567$ & $-0.0026$ & $+0.0009$ & $+0.0002$ & $-0.0029$ & $\mathbf{+0.0015}$ & $+0.0015$ \\
 & IS$_{95}$ & $0.1838$ & $-0.0057$ & $\mathbf{+0.0040^{**}}$ & $+0.0035$ & $-0.0056$ & $+0.0018$ & $+0.0017$ \\
 & CSL$_{L}$ & $-0.0681$ & $-0.0034$ & $+0.0126$ & $\mathbf{+0.0141^{*}}$ & $-0.0117$ & $+0.0047$ & $+0.0046$ \\
 & CSL$_{R}$ & $0.0614$ & $-0.0033$ & $+0.0020$ & $\mathbf{+0.0024}$ & $-0.0053$ & $+0.0002$ & $+0.0002$ \\
\addlinespace
CAD & KS & $0.079$ & $0.051$ & $0.041$ & $\mathbf{0.037}$ & $0.041$ & $0.050$ & $0.050$ \\
 & LogS & $2.1847$ & $-0.0091$ & $\mathbf{+0.0045}$ & $-0.0072$ & $-0.0036$ & $-0.0019$ & $-0.0020$ \\
 & CRPS & $0.0149$ & $-0.0001$ & $\mathbf{-0.0000}$ & $-0.0000$ & $-0.0000$ & $-0.0001$ & $-0.0001$ \\
 & IS$_{90}$ & $0.1208$ & $+0.0012$ & $+0.0008$ & $\mathbf{+0.0020^{*}}$ & $+0.0013$ & $-0.0009$ & $-0.0009$ \\
 & IS$_{95}$ & $0.1432$ & $-0.0003$ & $-0.0000$ & $\mathbf{+0.0012}$ & $-0.0001$ & $-0.0023$ & $-0.0022$ \\
 & CSL$_{L}$ & $0.0147$ & $\mathbf{+0.0079}$ & $+0.0071$ & $+0.0032$ & $-0.0008$ & $-0.0090$ & $-0.0091$ \\
 & CSL$_{R}$ & $0.1521$ & $-0.0061$ & $+0.0007$ & $+0.0047$ & $-0.0040$ & $\mathbf{+0.0055}$ & $+0.0055$ \\
\addlinespace
EUR & KS & $0.108^{\dagger}$ & $0.060$ & $0.052$ & $0.054$ & $0.068$ & $\mathbf{0.050}$ & $0.060$ \\
 & LogS & $2.1381$ & $+0.0221$ & $+0.0301^{*}$ & $+0.0303$ & $+0.0057$ & $\mathbf{+0.0513^{**}}$ & $+0.0441^{**}$ \\
 & CRPS & $0.0156$ & $\mathbf{+0.0003^{*}}$ & $+0.0001$ & $+0.0000$ & $+0.0002^{*}$ & $+0.0003^{*}$ & $+0.0002$ \\
 & IS$_{90}$ & $0.1304$ & $-0.0010$ & $+0.0051^{**}$ & $\mathbf{+0.0054^{**}}$ & $-0.0011$ & $+0.0053$ & $+0.0054$ \\
 & IS$_{95}$ & $0.1523$ & $-0.0100$ & $+0.0051$ & $+0.0055^{*}$ & $-0.0107$ & $\mathbf{+0.0085}$ & $+0.0081$ \\
 & CSL$_{L}$ & $-0.0032$ & $-0.0085$ & $+0.0198$ & $+0.0152$ & $-0.0144$ & $\mathbf{+0.0308^{**}}$ & $+0.0250^{*}$ \\
 & CSL$_{R}$ & $0.1095$ & $\mathbf{+0.0196}$ & $+0.0135$ & $+0.0128$ & $+0.0071$ & $+0.0136$ & $+0.0106$ \\
\addlinespace
JPY & KS & $0.088$ & $0.081$ & $0.061$ & $0.065$ & $0.065$ & $0.063$ & $\mathbf{0.053}$ \\
 & LogS & $2.1799$ & $+0.0015$ & $+0.0247$ & $+0.0238$ & $+0.0190$ & $\mathbf{+0.0507^{***}}$ & $+0.0498^{**}$ \\
 & CRPS & $0.0147$ & $-0.0000$ & $+0.0001$ & $+0.0001$ & $+0.0001$ & $+0.0002$ & $\mathbf{+0.0003^{*}}$ \\
 & IS$_{90}$ & $0.1245$ & $+0.0022$ & $+0.0033$ & $+0.0034$ & $+0.0034$ & $+0.0085^{***}$ & $\mathbf{+0.0093^{***}}$ \\
 & IS$_{95}$ & $0.1452$ & $-0.0007$ & $+0.0004$ & $+0.0004$ & $+0.0006$ & $\mathbf{+0.0090^{*}}$ & $+0.0088$ \\
 & CSL$_{L}$ & $0.0389$ & $-0.0011$ & $+0.0007$ & $+0.0007$ & $-0.0004$ & $+0.0158$ & $\mathbf{+0.0184}$ \\
 & CSL$_{R}$ & $0.0865$ & $-0.0051$ & $+0.0127$ & $+0.0131$ & $+0.0075$ & $\mathbf{+0.0209}$ & $+0.0194$ \\
\addlinespace
NZD & KS & $0.067$ & $\mathbf{0.038}$ & $0.044$ & $0.042$ & $0.063$ & $0.053$ & $0.053$ \\
 & LogS & $1.8610$ & $-0.0164$ & $-0.0148$ & $-0.0378^{***}$ & $\mathbf{-0.0000}$ & $-0.0006$ & $-0.0007$ \\
 & CRPS & $0.0210$ & $\mathbf{+0.0000}$ & $-0.0000$ & $-0.0001$ & $+0.0000$ & $-0.0001$ & $-0.0001$ \\
 & IS$_{90}$ & $0.1660$ & $+0.0001$ & $\mathbf{+0.0023}$ & $-0.0011$ & $+0.0015$ & $+0.0008$ & $+0.0008$ \\
 & IS$_{95}$ & $0.1946$ & $-0.0004$ & $+0.0041^{**}$ & $+0.0001$ & $-0.0008$ & $\mathbf{+0.0053}$ & $+0.0053$ \\
 & CSL$_{L}$ & $-0.0410$ & $+0.0050$ & $+0.0053$ & $\mathbf{+0.0075}$ & $-0.0028$ & $+0.0026$ & $+0.0026$ \\
 & CSL$_{R}$ & $0.1024$ & $-0.0065$ & $-0.0064$ & $-0.0102^{*}$ & $+0.0009$ & $\mathbf{+0.0048}$ & $+0.0048$ \\
\addlinespace
NOK & KS & $0.066$ & $0.052$ & $0.037$ & $\mathbf{0.036}$ & $0.048$ & $0.043$ & $0.050$ \\
 & LogS & $1.9447$ & $-0.0079$ & $\mathbf{+0.0062}$ & $-0.0066$ & $-0.0243$ & $-0.0081$ & $-0.0162$ \\
 & CRPS & $0.0191$ & $\mathbf{-0.0000}$ & $-0.0001$ & $+0.0000$ & $+0.0000$ & $-0.0000$ & $-0.0003$ \\
 & IS$_{90}$ & $0.1491$ & $-0.0018$ & $+0.0022$ & $\mathbf{+0.0028}$ & $-0.0039$ & $-0.0030$ & $-0.0023$ \\
 & IS$_{95}$ & $0.1727$ & $-0.0058$ & $+0.0018$ & $\mathbf{+0.0050}$ & $-0.0112$ & $-0.0062$ & $+0.0005$ \\
 & CSL$_{L}$ & $-0.0032$ & $-0.0113$ & $\mathbf{+0.0109}$ & $+0.0032$ & $-0.0192$ & $+0.0052$ & $-0.0107$ \\
 & CSL$_{R}$ & $0.0432$ & $-0.0091$ & $\mathbf{-0.0039}$ & $-0.0043$ & $-0.0115$ & $-0.0176$ & $-0.0138$ \\
\addlinespace
SEK & KS & $0.069$ & $0.057$ & $0.046$ & $0.045$ & $\mathbf{0.042}$ & $0.047$ & $0.062$ \\
 & LogS & $1.9755$ & $-0.0437$ & $-0.0032$ & $-0.0040$ & $-0.0087$ & $\mathbf{+0.0166}$ & $-0.0263$ \\
 & CRPS & $0.0186$ & $-0.0000$ & $-0.0001$ & $-0.0001$ & $+0.0000$ & $\mathbf{+0.0001}$ & $-0.0008^{*}$ \\
 & IS$_{90}$ & $0.1417$ & $-0.0044$ & $+0.0032$ & $+0.0035^{*}$ & $-0.0036$ & $\mathbf{+0.0040}$ & $-0.0085$ \\
 & IS$_{95}$ & $0.1637$ & $-0.0036$ & $+0.0068^{***}$ & $+0.0067^{***}$ & $-0.0065$ & $\mathbf{+0.0089^{*}}$ & $-0.0103$ \\
 & CSL$_{L}$ & $0.0267$ & $-0.0031$ & $+0.0061$ & $+0.0058$ & $-0.0045$ & $\mathbf{+0.0184}$ & $-0.0087$ \\
 & CSL$_{R}$ & $0.0616$ & $-0.0078$ & $-0.0128$ & $-0.0114$ & $-0.0062$ & $\mathbf{-0.0009}$ & $-0.0249$ \\
\addlinespace
CHF & KS & $0.103^{\dagger}$ & $0.055$ & $0.061$ & $0.057$ & $0.051$ & $\mathbf{0.050}$ & $0.050$ \\
 & LogS & $2.0843$ & $+0.0256$ & $+0.0136$ & $+0.0149$ & $+0.0075$ & $\mathbf{+0.0430^{**}}$ & $+0.0429^{**}$ \\
 & CRPS & $0.0163$ & $\mathbf{+0.0003^{*}}$ & $+0.0000$ & $+0.0001$ & $+0.0002$ & $+0.0001$ & $+0.0001$ \\
 & IS$_{90}$ & $0.1378$ & $+0.0019$ & $+0.0045^{**}$ & $+0.0046^{***}$ & $+0.0008$ & $\mathbf{+0.0060}$ & $+0.0060$ \\
 & IS$_{95}$ & $0.1652$ & $-0.0072$ & $+0.0055^{***}$ & $+0.0056^{***}$ & $-0.0021$ & $\mathbf{+0.0089}$ & $+0.0088$ \\
 & CSL$_{L}$ & $0.0745$ & $+0.0125$ & $+0.0112$ & $+0.0120$ & $+0.0103$ & $\mathbf{+0.0295^{***}}$ & $+0.0294^{***}$ \\
 & CSL$_{R}$ & $0.0331$ & $\mathbf{+0.0082}$ & $+0.0023$ & $+0.0026$ & $-0.0157$ & $+0.0034$ & $+0.0033$ \\
\addlinespace
GBP & KS & $0.084$ & $0.050$ & $0.045$ & $0.037$ & $0.039$ & $0.032$ & $\mathbf{0.030}$ \\
 & LogS & $2.2201$ & $-0.0111$ & $+0.0087$ & $+0.0055$ & $-0.0314$ & $+0.0177$ & $\mathbf{+0.0192}$ \\
 & CRPS & $0.0142$ & $+0.0000$ & $+0.0000$ & $+0.0000$ & $\mathbf{+0.0001}$ & $+0.0001$ & $+0.0001$ \\
 & IS$_{90}$ & $0.1175$ & $-0.0024$ & $\mathbf{+0.0040^{**}}$ & $+0.0032$ & $+0.0019$ & $+0.0034$ & $+0.0037$ \\
 & IS$_{95}$ & $0.1414$ & $-0.0045$ & $+0.0011$ & $-0.0004$ & $-0.0031$ & $+0.0041$ & $\mathbf{+0.0043}$ \\
 & CSL$_{L}$ & $-0.0011$ & $-0.0167$ & $+0.0062$ & $\mathbf{+0.0074}$ & $-0.0426$ & $+0.0028$ & $+0.0038$ \\
 & CSL$_{R}$ & $0.0692$ & $\mathbf{+0.0126}$ & $+0.0034$ & $+0.0063$ & $+0.0090$ & $+0.0118$ & $+0.0124$ \\
\bottomrule\end{tabular}
\caption{Exchange rates, BPS on the BPS evaluation window. Conventions as in Table~\ref{tab:infl-scores}.}
\label{tab:fx-scores-bps}
\end{table}

\newpage
\section{Component models for the exchange-rate application}
\label{app:tvpvar}
Each model is a time-varying parameter Bayesian vector autoregression (TVP-BVAR) in the form of \citet{Beckmann2020},
\begin{equation}
  \begin{aligned}
    y_t &= X_t \beta_t + \varepsilon_t, & \varepsilon_t &\sim \mathcal{N}(0, \Sigma_t),\\
    \beta_{t+1} &= \beta_t + \eta_t, & \eta_t &\sim \mathcal{N}(0, Q_t),
  \end{aligned}
  \label{eq:tvpvar}
\end{equation}
where $y_t$ is an $m \times 1$ vector of observed variables. The design matrix $X_t$ contains an intercept, $p$ lags of the endogenous variables (own lags and cross-lags) and one lag of the exogenous variables. With $n_x$ asset-specific and $n_{xx}$ non-asset-specific exogenous variables, $\beta_t$ has $k = m(1 + pm + n_x + n_{xx})$ elements. The coefficients start from a Minnesota-type prior, $\beta_0 \sim \mathcal{N}(0, \Omega_0)$, whose prior covariance $\Omega_0$ is governed by the shrinkage parameters $\gamma_1, \gamma_2, \dots$ described below. Smaller values imply stronger shrinkage towards zero.

The model is estimated by forgetting-factor recursions, which avoid posterior simulation. The state covariance $Q_t$ is implied by a forgetting factor $\lambda \in (0, 1]$: values below one let the coefficients drift, and $\lambda = 1$ gives $Q_t = 0$, so that $\beta_t$ is constant over time and \eqref{eq:tvpvar} reduces to a recursively updated constant-parameter VAR. The measurement covariance $\Sigma_t$ is estimated by an exponentially weighted moving average with decay $\kappa$. Further details are given in \citet{Beckmann2020} and Appendix~D of \citet{Bernaciak2024}.

The pool consists of every combination of the hyper-parameter values in Table~\ref{tab:tvpvar-grid}. We use the small pool of \citet[Section~4.2]{Bernaciak2024}: every model includes the UIP fundamental and has time-constant coefficients. 

\begin{table}[H]
\centering
\footnotesize
\onehalfspacing
\begin{tabular}{l l c}
\toprule
Parameter & Grid & Size \\
\midrule
Number of lags $p$ & $\mathcal{S}_p = \{6\}$ & 1 \\
Coefficient forgetting factor $\lambda$ & $\mathcal{S}_\lambda = \{1\}$ & 1 \\
EWMA decay of $\Sigma_t$, $\kappa$ & $\mathcal{S}_\kappa = \{0.97\}$ & 1 \\
Discount factor $\alpha$ & $\mathcal{S}_\alpha = \{1\}$ & 1 \\
Intercept shrinkage $\gamma_1$ & $\mathcal{S}_{\gamma_1} = \{0, 10\}$ & 2 \\
Own-lag shrinkage $\gamma_2$ & $\mathcal{S}_{\gamma_2} = \{0, 0.1, 0.5, 0.9\}$ & 4 \\
Cross-lag shrinkage $\gamma_3$ & $\mathcal{S}_{\gamma_3} = \{0, 0.1, 0.5, 0.9\}$ & 4 \\
UIP inclusion $\gamma_4$ & $\mathcal{S}_{\gamma_4} = \{1\}$ & 1 \\
\midrule
Number of models & & 32 \\
\bottomrule
\end{tabular}
\caption{Hyper-parameter grid of the TVP-BVAR component models, following the small pool of \citet{Bernaciak2024}.}
\label{tab:tvpvar-grid}
\end{table}

\section{Exchange rate results}
\begin{table}[H]\centering\tiny
\begin{tabular}{llccccccc}\toprule
Currency & Metric & Raw pool & Static-RQS & RQS-GAS & RQS-Con & Static-Beta & Beta-GAS & Beta-Con \\ \midrule
AUD & KS & $0.043$ & $0.050$ & $0.037$ & $\mathbf{0.031}$ & $0.056$ & $0.045$ & $0.049$ \\
 & LogS & $1.8971$ & $\mathbf{+0.0011}$ & $-0.0225^{***}$ & $-0.0174^{*}$ & $-0.0015$ & $-0.0065$ & $-0.0136$ \\
 & CRPS & $0.0201$ & $-0.0001^{*}$ & $-0.0001$ & $\mathbf{-0.0000}$ & $-0.0001$ & $-0.0002$ & $-0.0003^{*}$ \\
 & IS$_{90}$ & $0.1528$ & $+0.0012$ & $-0.0009$ & $+0.0008$ & $\mathbf{+0.0013}$ & $-0.0025$ & $-0.0034$ \\
 & IS$_{95}$ & $0.1755$ & $\mathbf{+0.0027}$ & $-0.0000$ & $+0.0012$ & $+0.0015$ & $-0.0032$ & $-0.0059$ \\
 & CSL$_{L}$ & $0.0168$ & $-0.0051$ & $-0.0047$ & $\mathbf{+0.0009}$ & $-0.0032$ & $-0.0047$ & $-0.0074$ \\
 & CSL$_{R}$ & $0.1068$ & $\mathbf{+0.0029}$ & $-0.0095^{*}$ & $-0.0009$ & $-0.0035$ & $-0.0182^{**}$ & $-0.0219^{**}$ \\
\addlinespace
CAD & KS & $0.054$ & $0.063$ & $0.042$ & $0.039$ & $0.055$ & $0.044$ & $\mathbf{0.034}$ \\
 & LogS & $2.2396$ & $-0.0425^{**}$ & $-0.0172$ & $-0.0165$ & $-0.0482^{**}$ & $-0.0106$ & $\mathbf{+0.0020}$ \\
 & CRPS & $0.0139$ & $-0.0003^{**}$ & $-0.0002$ & $\mathbf{-0.0001}$ & $-0.0003^{**}$ & $-0.0003^{**}$ & $-0.0001$ \\
 & IS$_{90}$ & $0.1103$ & $-0.0008$ & $+0.0006$ & $+0.0004$ & $-0.0026$ & $+0.0001$ & $\mathbf{+0.0014}$ \\
 & IS$_{95}$ & $0.1317$ & $-0.0012$ & $-0.0001$ & $+0.0000$ & $-0.0022$ & $-0.0006$ & $\mathbf{+0.0011}$ \\
 & CSL$_{L}$ & $0.0365$ & $-0.0160^{***}$ & $-0.0153^{***}$ & $-0.0159^{***}$ & $-0.0100^{**}$ & $-0.0127$ & $\mathbf{-0.0099}$ \\
 & CSL$_{R}$ & $0.1076$ & $-0.0226^{*}$ & $-0.0133^{**}$ & $-0.0138^{**}$ & $-0.0396^{*}$ & $-0.0146^{*}$ & $\mathbf{-0.0090}$ \\
\addlinespace
EUR & KS & $0.066$ & $0.042$ & $0.042$ & $0.042$ & $0.057$ & $0.040$ & $\mathbf{0.037}$ \\
 & LogS & $2.1573$ & $-0.0211$ & $-0.0165^{*}$ & $-0.0368^{***}$ & $-0.0022$ & $\mathbf{+0.0005}$ & $-0.0060$ \\
 & CRPS & $0.0157$ & $\mathbf{-0.0000}$ & $+0.0000$ & $-0.0003^{**}$ & $-0.0000$ & $-0.0001$ & $-0.0002$ \\
 & IS$_{90}$ & $0.1248$ & $-0.0021$ & $-0.0005$ & $\mathbf{+0.0001}$ & $-0.0015$ & $-0.0000$ & $-0.0013$ \\
 & IS$_{95}$ & $0.1458$ & $-0.0026$ & $-0.0017$ & $\mathbf{-0.0011}$ & $-0.0019$ & $-0.0017$ & $-0.0033$ \\
 & CSL$_{L}$ & $-0.0230$ & $+0.0039$ & $+0.0049$ & $-0.0004$ & $-0.0004$ & $\mathbf{+0.0120}$ & $+0.0082$ \\
 & CSL$_{R}$ & $0.0099$ & $-0.0054$ & $-0.0084^{*}$ & $-0.0137^{**}$ & $\mathbf{-0.0048}$ & $-0.0106$ & $-0.0138^{*}$ \\
\addlinespace
JPY & KS & $0.073$ & $0.054$ & $\mathbf{0.050}$ & $0.066$ & $0.068$ & $0.053$ & $0.078$ \\
 & LogS & $2.1074$ & $+0.0174$ & $+0.0035$ & $-0.0187$ & $+0.0121$ & $\mathbf{+0.0184}$ & $-0.0051$ \\
 & CRPS & $0.0164$ & $\mathbf{+0.0001^{**}}$ & $+0.0000$ & $-0.0002$ & $+0.0001$ & $+0.0001$ & $-0.0003$ \\
 & IS$_{90}$ & $0.1343$ & $+0.0016$ & $+0.0009$ & $+0.0010$ & $+0.0012$ & $\mathbf{+0.0027}$ & $-0.0010$ \\
 & IS$_{95}$ & $0.1603$ & $+0.0010$ & $+0.0013$ & $+0.0004$ & $+0.0013$ & $\mathbf{+0.0018}$ & $+0.0018$ \\
 & CSL$_{L}$ & $-0.0167$ & $\mathbf{+0.0076}$ & $+0.0041$ & $-0.0006$ & $+0.0046$ & $+0.0057$ & $-0.0168$ \\
 & CSL$_{R}$ & $0.0207$ & $\mathbf{+0.0058}$ & $-0.0062$ & $-0.0149$ & $+0.0021$ & $+0.0003$ & $-0.0010$ \\
\addlinespace
NZD & KS & $0.043$ & $0.064$ & $\mathbf{0.033}$ & $0.034$ & $0.051$ & $0.041$ & $0.045$ \\
 & LogS & $1.8706$ & $-0.0242^{*}$ & $-0.0098$ & $-0.0248^{***}$ & $\mathbf{-0.0048}$ & $-0.0162$ & $-0.0165$ \\
 & CRPS & $0.0211$ & $\mathbf{-0.0001^{*}}$ & $-0.0002$ & $-0.0002$ & $-0.0001$ & $-0.0003^{*}$ & $-0.0003$ \\
 & IS$_{90}$ & $0.1645$ & $-0.0001$ & $-0.0022$ & $-0.0020$ & $\mathbf{+0.0000}$ & $-0.0031$ & $-0.0009$ \\
 & IS$_{95}$ & $0.1898$ & $-0.0017$ & $-0.0018$ & $\mathbf{-0.0001}$ & $-0.0009$ & $-0.0054$ & $-0.0065$ \\
 & CSL$_{L}$ & $-0.0175$ & $+0.0028$ & $\mathbf{+0.0078}$ & $+0.0003$ & $-0.0015$ & $-0.0023$ & $-0.0005$ \\
 & CSL$_{R}$ & $0.0231$ & $-0.0066$ & $-0.0106^{**}$ & $-0.0139^{***}$ & $\mathbf{-0.0006}$ & $-0.0120$ & $-0.0049$ \\
\addlinespace
NOK & KS & $0.054$ & $\mathbf{0.040}$ & $0.049$ & $0.041$ & $0.048$ & $0.048$ & $0.048$ \\
 & LogS & $2.0006$ & $-0.0436^{***}$ & $-0.0174^{*}$ & $-0.0394^{**}$ & $-0.0247^{**}$ & $-0.0120$ & $\mathbf{-0.0082}$ \\
 & CRPS & $0.0181$ & $-0.0001$ & $\mathbf{+0.0000}$ & $-0.0003$ & $-0.0001$ & $-0.0003^{*}$ & $-0.0001$ \\
 & IS$_{90}$ & $0.1412$ & $-0.0044^{*}$ & $\mathbf{+0.0017}$ & $-0.0022$ & $-0.0037^{*}$ & $-0.0024$ & $-0.0030$ \\
 & IS$_{95}$ & $0.1642$ & $-0.0069^{**}$ & $\mathbf{+0.0010}$ & $-0.0015$ & $-0.0081^{**}$ & $-0.0038$ & $-0.0058$ \\
 & CSL$_{L}$ & $0.0095$ & $+0.0010$ & $+0.0027$ & $-0.0047$ & $-0.0038$ & $+0.0060$ & $\mathbf{+0.0126}$ \\
 & CSL$_{R}$ & $0.0151$ & $-0.0259^{**}$ & $\mathbf{-0.0028}$ & $-0.0102$ & $-0.0200^{**}$ & $-0.0151^{**}$ & $-0.0195^{**}$ \\
\addlinespace
SEK & KS & $0.055$ & $0.045$ & $\mathbf{0.025}$ & $0.032$ & $0.046$ & $0.037$ & $0.037$ \\
 & LogS & $2.0373$ & $-0.0333^{**}$ & $-0.0123$ & $-0.0119$ & $-0.0141$ & $\mathbf{-0.0071}$ & $-0.0082$ \\
 & CRPS & $0.0176$ & $\mathbf{-0.0001}$ & $-0.0002$ & $-0.0002$ & $-0.0001$ & $-0.0002$ & $-0.0002$ \\
 & IS$_{90}$ & $0.1322$ & $-0.0059^{**}$ & $-0.0011$ & $-0.0011$ & $-0.0026$ & $\mathbf{-0.0007}$ & $-0.0010$ \\
 & IS$_{95}$ & $0.1519$ & $-0.0075^{*}$ & $+0.0005$ & $+0.0018$ & $-0.0032$ & $\mathbf{+0.0021}$ & $+0.0012$ \\
 & CSL$_{L}$ & $0.0686$ & $+0.0027$ & $-0.0019$ & $-0.0016$ & $+0.0040$ & $+0.0052$ & $\mathbf{+0.0058}$ \\
 & CSL$_{R}$ & $0.0454$ & $-0.0286^{**}$ & $-0.0122^{*}$ & $-0.0091$ & $-0.0154^{*}$ & $\mathbf{-0.0070}$ & $-0.0076$ \\
\addlinespace
CHF & KS & $0.067$ & $0.048$ & $0.042$ & $0.048$ & $0.048$ & $\mathbf{0.036}$ & $0.042$ \\
 & LogS & $2.0951$ & $-0.0021$ & $-0.0108$ & $-0.0259^{*}$ & $\mathbf{+0.0081^{*}}$ & $-0.0029$ & $-0.0245$ \\
 & CRPS & $0.0165$ & $\mathbf{+0.0001^{*}}$ & $-0.0002$ & $-0.0003^{**}$ & $+0.0001$ & $-0.0002^{*}$ & $-0.0004^{**}$ \\
 & IS$_{90}$ & $0.1320$ & $\mathbf{+0.0015^{*}}$ & $+0.0004$ & $-0.0003$ & $+0.0015^{*}$ & $+0.0006$ & $-0.0025$ \\
 & IS$_{95}$ & $0.1590$ & $\mathbf{+0.0027^{**}}$ & $+0.0004$ & $+0.0001$ & $+0.0024^{*}$ & $-0.0004$ & $-0.0049$ \\
 & CSL$_{L}$ & $0.0357$ & $\mathbf{+0.0112^{***}}$ & $+0.0065$ & $-0.0002$ & $+0.0106^{***}$ & $+0.0097$ & $+0.0070$ \\
 & CSL$_{R}$ & $-0.0343$ & $-0.0019$ & $-0.0046$ & $\mathbf{-0.0000}$ & $-0.0064^{***}$ & $-0.0091$ & $-0.0247^{**}$ \\
\addlinespace
GBP & KS & $0.062$ & $0.037$ & $0.034$ & $0.032$ & $0.037$ & $\mathbf{0.031}$ & $0.035$ \\
 & LogS & $2.2498$ & $\mathbf{+0.0089}$ & $-0.0166^{*}$ & $-0.0109$ & $-0.0010$ & $-0.0089$ & $-0.0070$ \\
 & CRPS & $0.0134$ & $\mathbf{+0.0000}$ & $-0.0002^{***}$ & $-0.0002^{**}$ & $-0.0000$ & $-0.0002^{***}$ & $-0.0002^{***}$ \\
 & IS$_{90}$ & $0.1104$ & $\mathbf{+0.0019}$ & $+0.0007$ & $+0.0008$ & $+0.0014$ & $+0.0017$ & $+0.0019$ \\
 & IS$_{95}$ & $0.1287$ & $-0.0019$ & $-0.0005$ & $-0.0012$ & $-0.0013$ & $-0.0007$ & $\mathbf{-0.0002}$ \\
 & CSL$_{L}$ & $0.0592$ & $\mathbf{+0.0030}$ & $-0.0088$ & $-0.0064$ & $-0.0041$ & $-0.0050$ & $-0.0042$ \\
 & CSL$_{R}$ & $0.0784$ & $-0.0074$ & $-0.0085^{*}$ & $-0.0064$ & $\mathbf{-0.0028}$ & $-0.0065$ & $-0.0055$ \\
\bottomrule\end{tabular}
\caption{Exchange rates, $\alpha=0.9$, full evaluation window. Conventions as in Table~\ref{tab:infl-scores}.}
\label{tab:fx-scores}
\end{table}

\newpage

\newpage
\bibliographystyle{apalike}
\bibliography{references}

\end{document}